\documentclass[twocolumn]{aastex61}
\pdfoutput=1

\usepackage{morefloats}
\usepackage{epstopdf}
\usepackage{xcolor, graphicx}
\usepackage{amssymb}
\usepackage{amsmath}
\usepackage{natbib}
\usepackage{multirow}
\usepackage{gensymb}
\usepackage[latin10]{inputenc}

\begin{document}


\title{GEOMETRIC SUPPORT FOR DARK MATTER BY AN UNALIGNED EINSTEIN RING IN ABELL 3827}

\author{Mandy C. Chen}\affiliation{Department of Astronomy and Astrophysics, The University of Chicago, Chicago, IL 60637, USA}\affiliation{Department of Physics, The University of Hong Kong, Pokfulam Road, Hong Kong}
\author{Tom Broadhurst}\affiliation{Department of Theoretical Physics, University of the Basque Country UPV/EHU, Bilbao, Spain}\affiliation{IKERBASQUE, Basque Foundation for Science, Bilbao, Spain}\affiliation{Donostia International Physics Center (DIPC), 20018 Donostia, Spain}
\author{Jeremy Lim}\affiliation{Department of Physics, The University of Hong Kong, Pokfulam Road, Hong Kong}
\author{Sandor M. Molnar}\affiliation{Institute of Astronomy and Astrophysics, Academia Sinica, P. O. Box
23-141, Taipei 10617, Taiwan}
\author{Jose M.~Diego}\affiliation{IFCA, Instituto de F\'isica de Cantabria (UC-CSIC), Av.~de Los Castros s/n, 39005 Santander, Spain}
\author{Masamune Oguri}\affiliation{Research Center for the Early Universe, University of Tokyo, 7-3-1 Hongo, Bunkyo-ku, Tokyo 113-0033, Japan}\affiliation{Department of Physics, University of Tokyo, 7-3-1 Hongo, Bunkyo-ku, Tokyo 113-0033, Japan}\affiliation{Kavli Institute for the Physics and Mathematics of the Universe (Kavli IPMU, WPI), University of Tokyo, Chiba 277-8583, Japan}
\author{Lilian L. Lee}\affiliation{Department of Physics, The University of Hong Kong, Pokfulam Road, Hong Kong}

\correspondingauthor{Mandy C. Chen}
\email{mandychen@astro.uchicago.edu}


\begin{abstract}

The non-detection of dark matter (DM) particles in increasingly stringent laboratory searches has encouraged alternative gravity theories where gravity is sourced only from visible matter. Here, we consider whether such theories can pass a two-dimensional test posed by gravitational lensing -- to reproduce a particularly detailed Einstein ring in the core of the galaxy cluster Abell 3827. We find that when we require the lensing mass distribution to strictly follow the shape (ellipticity and position angle) of the light distribution of cluster member galaxies, intracluster stars, and the X-ray emitting intracluster medium, we cannot reproduce the Einstein ring, despite allowing the mass-to-light ratios of these visible components to freely vary with radius to mimic alternative gravity theories. Alternatively, we show that the detailed features of the Einstein ring are accurately reproduced by allowing a smooth, freely oriented DM halo in the lens model, with relatively small contributions from the visible components at a level consistent with their observed brightnesses. This dominant DM component is constrained to have the same orientation as the light from the intracluster stars, indicating that the intracluster stars trace the gravitational potential of this component. The Einstein ring of Abell 3827 therefore presents a new challenge for alternative gravity theories: not only must such theories find agreement between the total lensing mass and visible mass, but they must also find agreement between the projected sky distribution of the lensing mass and that of the visible matter, a more stringent test than has hitherto been posed by lensing data.

\end{abstract}




\section{Introduction}

In the framework of general relativity (GR), predominantly invisible matter -- referred to as dark matter (DM) -- is required to explain galaxy rotation speeds, dynamics of galaxy clusters, gravitational lensing by galaxies and galaxy clusters, baryon acoustic oscillations, and anisotropies in the cosmic microwave background \citep[e.g.][]{Zwicky1933,Dodelson2003,Salucci2019}. Furthermore, observations of galaxy clusters such as the Bullet cluster \citep{Markevitch2004} as well as other colliding clusters \citep{MolnarBroadhurst2017} require DM to be predominantly non-relativistic and lacking any interaction with all known particles and with itself other than through gravity.   Having a cosmological mass density that is determined to be several times greater than that of familiar particles in the Standard Model of particle physics \citep[e.g.][]{Planck2018}, accepting GR therefore comes at the price of requiring fundamentally new physics.  

Despite increasingly stringent laboratory searches, no evidence has emerged for the favored heavy particle interpretation for DM \citep[e.g.][]{Xenon1T2017}.  This failure to directly detect DM warrants serious consideration of alternative gravity theories such as modified Newtonian dynamics \citep[MOND; e.g.][]{Milgrom1983,Bekenstein2011}, scale-invariant gravity \citep{Maeder2017,Maeder2019}, and emergent gravity \citep{Verlinde2017}.  Such theories of gravity, however, face the challenge of satisfying a wide range of strict tests that have already been passed by GR \citep[e.g.][]{Clowe2004,Hees2017}.  Among the most recent and well known of these tests are gravitational waves, detected first by the LIGO observatory but now also Virgo observatory, which are well described by merging black holes in binary systems obeying GR \citep[e.g.][]{LIGO2016}.   Moreover, in the strong-field regime, GR accounts for the observed ring of lensed radio emission around the event horizon of a supermassive black hole in the galaxy M87 \citep{EHT2019}.


In the context of this debate, we analyze the exceptionally complete and detailed Einstein ring around the center of the galaxy cluster Abell 3827 (hereafter A3827). This Einstein ring comprises multiply-lensed images of a background spiral galaxy having numerous resolved features, which therefore allow for detailed modeling of the lensing mass distribution.  The visible matter in and around the Einstein ring comprises four bright and dominant cluster member galaxies along with a dimmer member galaxy within this ring, intracluster light that extends well beyond the ring, and even more extended and nearly circular X-ray emission that is centered on the ring.  We carefully consider whether the multiply-lensed images can be generated by gravity that is sourced based solely upon the projected sky distribution of the aforementioned visible matter.
Any necessary additional component having a projected sky distribution different from that of the visible matter would then imply the presence of invisible matter that contributes in a measurable way to the bending of light by the cluster.


Previous lens modeling work on A3827 \citep{Carrasco2010,Williams2011,Massey1,Massey2} has utilized the special configuration of this system to investigate the nature of DM. Possible evidence for self-interacting DM was identified in one of the bright central galaxies \citep{Massey1,Taylor2017}. New data from ALMA, however, provided an unambiguous identification of multiple images that shows no evidence for DM self-interaction \citep{Massey2}. These studies have demonstrated the unique advantage of A3827 for understanding DM properties. 

In this paper, we first present the data used for analysis in Section~\ref{data}, including optical imaging data by {\it Hubble} and X-ray data by {\it Chandra}. Then we describe our lens modeling process in Section~\ref{lens_modeling}, where we construct both free-form and parametric lens models for A3827, and present key properties of these models. In Section~\ref{sec:discussion} we discuss the implications of our lens models for alternative theories of gravity and the necessity of DM, and finally in Section~\ref{sec:conclusion} we present a concise summary and our conclusions. Throughout this paper, we adopt a Hubble constant of $H_0=70$ km/s/Mpc, $\Omega_\mathrm{M}=0.3$ and $\Omega_\Lambda = 0.7$ when deriving distances, masses and luminosities. All magnitudes quoted are AB magnitudes. At the redshift of the cluster, $1\arcsec$ corresponds to $1.83$ kpc.

\section{Data analysis}\label{data}
\subsection{\textit{HST}}\label{section:HST_data}
\begin{figure*}[tp!]
\centering
\graphicspath{{/Users/mandychen/A3827/ApJ/figures/}}
\includegraphics[width=\linewidth]{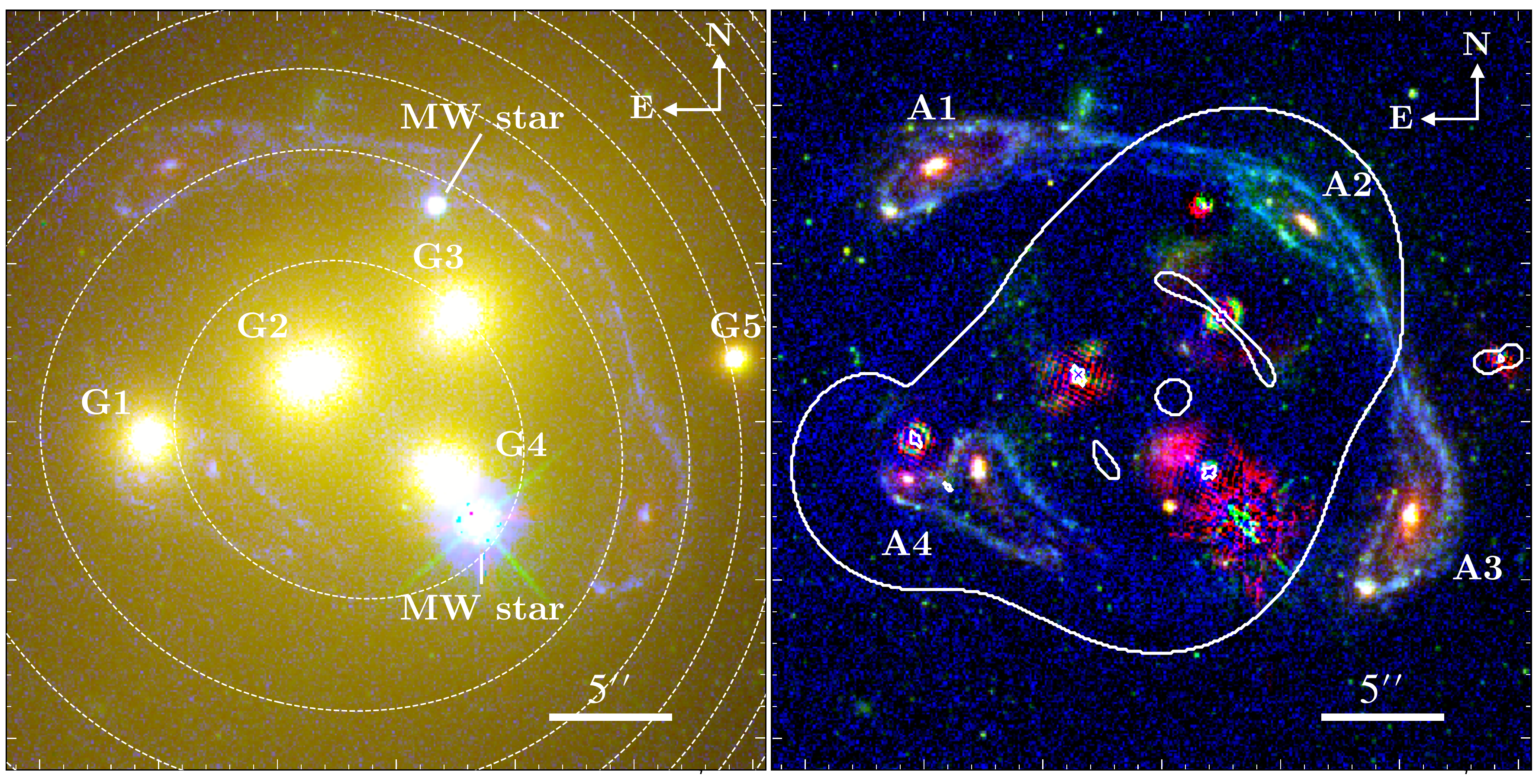}
\caption{\label{fig:hst_and_residual_map} {\it Left:} a color image of A3827 cluster core with a field of view (FOV) of $0.4\arcmin \times 0.4\arcmin$, composed with {\it HST} images from filters F336W (blue), F606W and F814W (green), and F160W (red). Labels indicate the four BCGs G1$-$G4, a compact cluster member galaxy G5, and two Milky Way (MW) stars. White dashed contours indicate the X-ray broadband emission obtained from {\it Chandra} (see \S\ref{Chandra_data}), separated according to the square root of emission intensity.  {\it Right:} a color image in the same FOV composed with residual images after subtracting the \texttt{IMFIT} galaxy light models from {\it HST} images (see \S\ref{section:IMFIT}). Labels show multiply-lensed images A1-A4. The galaxy bulge in image A4 is locally lensed into two images due to the lensing perturbation introduced by galaxy G1. The white contour shows the critical curve predicted by our best-fit {\it glafic} lens model (Model 3, see \S\ref{section:glafic}).  }
\end{figure*}

Deep imaging data of A3827 ($z=0.099$) taken by the \textit{Hubble Space Telescope} (\textit{HST}) of program GO-12817 \citep{Massey1,Massey2} comprise a total of 15,779s of exposure spanning four broadbands from UV (F336W), optical (F606W and F814W), to near-IR (F160W). From this we construct a multi-colour image of the cluster core, as shown in Figure~\ref{fig:hst_and_residual_map}. The most prominent feature is a nearly complete Einstein ring of $\sim10\arcsec$ in radius, encircling the four brightest cluster galaxies (BCGs) of comparable luminosities, from a background galaxy at $z = 1.24$.  

We name the four BCGs as G1--G4. Four multiply-lensed images A1-A4 comprising a red bulge surrounded by blue spiral arms can be immediately identified, while the galaxy bulge A4 (located very close to galaxy G1) is further split into two images owing to the local lensing perturbation induced by G1. There are two Milky Way (MW) stars projected close to the cluster core, both of which we have subtracted (see \S\ref{section:IMFIT}) from the image shown in Figure~\ref{fig:hst_and_residual_map} (right panel), along with a small compact cluster member galaxy (G5) located to the west of lensed images.  All of these components are indicated in Figure \ref{fig:hst_and_residual_map}.

\subsubsection{\texttt{IMFIT} model of galaxy light}\label{section:IMFIT}
To better reveal the lensed images, we fit the light distributions of the four BCGs (G1--G4), the two MW stars, the compact nearby galaxy G5, and the diffuse intracluster light (ICL) using \texttt{IMFIT} \citep{Erwin2015}\footnote{http://www.mpe.mpg.de/~erwin/code/IMFIT}. All of these sources are fitted simultaneously as they overlap on the sky. This fitting is performed in all filters except for the F336W UV band, where cluster member galaxies do not outshine the lensed images. We started by assigning to G1$-$G5 and the ICL a single S\'ersic function each. The two MW stars are fitted by simply scaling a point spread function (PSF) that is taken directly from a star in an isolated region in each filter. In Table~\ref{table:BCG_ellipticity}, we list the ellipticity ($\epsilon$) and position angles (PAs) of G1--G4 and the ICL from the best-fit models in the reddest F160W band; the best-fit parameters are comparable across all filters.  Except for G2, all three of the other BCGs are nearly circular in projection with low ellipticity values ($\epsilon<0.1$). G2 has an $\epsilon$ of 0.22, and the large halo describing the ICL has an $\epsilon$ of 0.29. 

\begin{deluxetable}{c | c c}[tp!]
\tablecaption{\label{table:BCG_ellipticity}Ellipticity ($\epsilon$) and position angles (PAs) of BCGs G1--G4 obtained from \texttt{IMFIT} modeling in the F160W band with one S\'ersic profile assigned to each galaxy. Ellipticity is defined as $\epsilon = 1-b/a$ where a and b are the major and minor axis, respectively. PAs are measured counter-clockwise from the North. Except for G2, all other three BCGs are nearly circular in projection with low $\epsilon$ values.}
\tablehead{\colhead{} & \colhead{$\epsilon$} & \colhead{PA}}
\startdata
G1 & 0.05 & 139.07 \\
G2 & 0.22 & 113.15 \\
G3 & 0.04 & 125.76 \\
G4 & 0.06 & 121.56 \\
ICL & 0.29 & 149.39 \\
\enddata
\end{deluxetable}

We also experimented with using two concentric S\'ersic functions, from which we obtain slightly improved models for G1--G4.  The ICL is also better fitted by two concentric S\'ersic functions in the F160W band, where it is detected to its furthest extent, while a single S\'ersic function provides a sufficiently good fit in the F814W and F606W band. The relatively dim galaxy G5 is well fitted by a single S\'ersic function. Note that for cases where two concentric S\'ersic functions provide a better fit, the dominating S\'ersic component (which is more spatially extended) has a similar $\epsilon$ and PA as the corresponding parameters obtained from a single S\'ersic fitting (difference in $\epsilon$ $<0.05$ and in PA $<10\degree$), indicating that the $\epsilon$ and PA derived from single S\'ersic models are robust in capturing the large-scale morphology of the individual galaxies.  

In each filter apart from F336W, we subtract the best-fit models derived in that filter for the foreground objects to minimize their contamination on the lensed images of the background galaxy. For the F336W image, we scaled the best-fit model of the F814W band and subtracted this scaled model from the data, with the scaling factor determined arbitrarily through trial and error. In Figure \ref{fig:hst_and_residual_map} (right-hand panel), we show a color image of the cluster core composed of model-subtracted data to highlight the details of the Einstein ring. 

To estimate the stellar mass contained in both the galaxies and the diffuse ICL, we obtain the stellar mass-to-light ($M/L$) ratio using the {\it Yggdrasil} model \citep{YGGmodel}, for which we assume a Salpeter IMF \citep{Salpeter1955} and a single burst of star formation at redshift $z=3$.  In this manner, we infer $(M/L)_{\odot,\mathrm{F606W}} = 4.8$.  We then estimate the stellar masses of G1$-$G5 and ICL based on their integrated flux from \texttt{IMFIT} models fitted simultaneously to all of these components within a central radius of $1.5\arcmin$ ($\sim 170$ kpc), which, based on a visual inspection, is where the ICL drops below the level of detectability. The results are listed in Table~\ref{table:stellar_mass}. Note that our stellar mass estimations are close to but do not completely agree with the values reported in Table 2 of \cite{Massey2}. The difference is possibly due to different choices of stellar population synthesis models, the radius for integrating the flux, and the method for fitting the two-dimensional light distribution.  These differences between our estimated stellar masses and those in \cite{Massey2} are smaller than $25\%$ and do not affect our subsequent analysis. 

\begin{figure}[tp!]
\centering
\includegraphics[width=\linewidth]{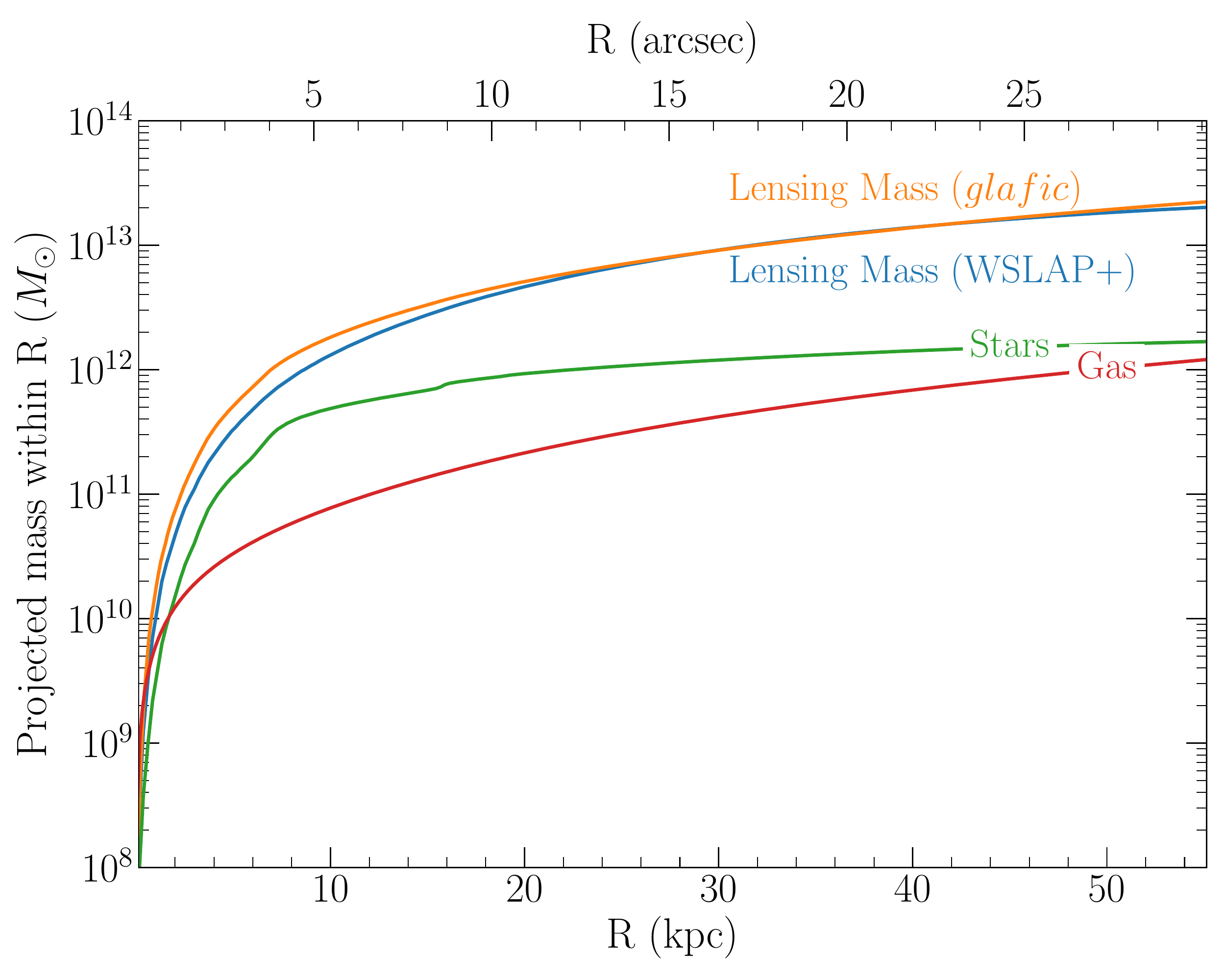}
\caption{\label{fig:1D_mass} Projected cumulative mass within radius R from the cluster center determined from the X-ray emission (RA$=330.4728\degree$, Dec$=-59.946383\degree$). The total lensing mass is based on WSLAP+ and {\it glafic} models (Model 1 and Model 3) as discussed in sections \ref{section:wslap} and \ref{section:glafic}, respectively. Stellar mass is converted from the F606W band flux based on the {\it Yggdrasil} model \citep{YGGmodel} with $(M_\star/L)_{\odot,\mathrm{F606W}} = 4.8$, as described in \S\ref{section:IMFIT}. We estimate the gas mass from {\it Chandra} data, as described in \S\ref{X-ray_gas_mass}. Within the Einstein ring at $\mathrm{R}\approx10\arcsec$, stellar and gas mass have a relatively small contribution to the total lensing mass. }
\end{figure}

To obtain the radial profile of the projected stellar mass, we also convert the \texttt{IMFIT} model from the F606W band into stellar mass. Figure \ref{fig:1D_mass} shows the projected mass within a given radius, {\it R}, of the cluster center as defined by the centroid of the cluster-scale X-ray emission (see \S\ref{X-ray_ellipticity}).  Note that the projected mass of the intracluster X-ray gas becomes increasingly larger than that of the stellar mass beyond the largest radius plotted in this figure.

\begin{deluxetable}{c | c c}[tp!]
\tablecaption{\label{table:stellar_mass} Total integrated magnitudes of galaxies G1-G5 and ICL within a radius of $1.5\arcmin$ ($\sim 170$ kpc) based on the best-fit \texttt{IMFIT} models. Corresponding stellar masses are converted based on the {\it Yggdrasil} model \citep{YGGmodel} with $(M_\star/L)_{\odot,\mathrm{F606W}} = 4.8$, as described in \S\ref{section:IMFIT}. }
\tablehead{\colhead{} & \colhead{M$_\mathrm{AB}$} & \colhead{$M_\star$ ($M_\odot$)}}
\startdata
G1 & 17.21 & $1.34\times 10^{11}$ \\
G2 & 16.51 & $2.55\times 10^{11}$ \\
G3 & 16.25 & $3.23\times 10^{11}$ \\
G4 & 16.75 & $2.04\times 10^{11}$ \\
G5 & 19.26 & $2.03\times 10^{10}$ \\
Intra-cluster light (ICL) & 14.61 & $1.47\times 10^{12}$ \\
\enddata
\end{deluxetable}


\subsubsection{Counter-image identification}\label{section:multiple_image}
Along the Einstein ring formed from the background lensed galaxy at $z=1.24$, we identify a total of 39 knots to use as constraints for lens modeling, corresponding to 9 distinctive compact features in the lensed galaxy. We list the coordinates of these knots in Table~\ref{image-constraints} and show their positions in Figure~\ref{fig:image_ID}. To determine counter-images, we first identify major features in the background lensed galaxy (e.g., its bulge) through visual inspection of the colors and morphologies of the knots.  With 21 major features (knots 1$-$4, see figure in Appendix) thus identified and serving as initial constraints, we then constructed a preliminary lens model to guide our identification of less obvious features (e.g., knot No. 9).  In this way, we reached our final set of constraints comprising 39 knots.

\begin{deluxetable}{c   c   c}
\tablecaption{\label{image-constraints}Coordinates of 39 multiply-lensed knots used as constraints for lens modeling, which correspond to 9 distinctive compact features in the background galaxy. Locations of these knots are shown in the Appendix.}
\tablehead{
\colhead{Knot ID} & \colhead{RA} (J2000) & \colhead{DEC} (J2000)}
\startdata
1.1 & 22:01:53.950 & -59:56:36.91 \\
1.2 & 22:01:52.376 & -59:56:38.83 \\ 
1.3 & 22:01:51.957 & -59:56:47.98 \\ 
1.4 & 22:01:53.768 & -59:56:46.55 \\
1.5 & 22:01:54.072 & -59:56:46.82 \\
1.6 & 22:01:52.899 & -59:56:42.73 \\ 
1.7 & 22:01:52.882 & -59:56:46.21 \\
2.1 & 22:01:54.131 & -59:56:38.44 \\ 
2.2 & 22:01:52.486 & -59:56:39.15 \\
2.3 & 22:01:52.131 & -59:56:50.37 \\ 
2.4 & 22:01:53.861 & -59:56:45.74 \\ 
2.5 & 22:01:54.103 & -59:56:47.47 \\ 
3.1 & 22:01:54.169 & -59:56:38.26 \\ 
3.2 & 22:01:52.443 & -59:56:39.43 \\ 
3.3 & 22:01:52.167 & -59:56:50.19 \\ 
3.5 & 22:01:54.019 & -59:56:47.76 \\ 
4.1 & 22:01:53.975 & -59:56:38.19 \\ 
4.2 & 22:01:52.611 & -59:56:38.32 \\
4.3 & 22:01:52.002 & -59:56:50.15 \\
4.4 & 22:01:53.808 & -59:56:45.48 \\
4.5 & 22:01:54.162 & -59:56:46.92 \\ 
5.1 & 22:01:53.646 & -59:56:36.87 \\
5.2 & 22:01:52.714 & -59:56:37.34 \\
5.3 & 22:01:51.820 & -59:56:48.69 \\
5.4 & 22:01:53.683 & -59:56:45.68 \\
6.1 & 22:01:53.655 & -59:56:35.74 \\
6.2 & 22:01:52.514 & -59:56:37.29 \\
6.3 & 22:01:51.796 & -59:56:46.86 \\
7.1 & 22:01:54.082 & -59:56:36.94 \\
7.2 & 22:01:52.281 & -59:56:40.08 \\ 
7.3 & 22:01:52.041 & -59:56:47.70 \\
7.4 & 22:01:53.446 & -59:56:49.20 \\ 
8.1 & 22:01:54.024 & -59:56:36.36 \\
8.3 & 22:01:51.967 & -59:56:46.28 \\
8.4 & 22:01:53.457 & -59:56:48.57 \\ 
9.1 & 22:01:53.925 & -59:56:35.80 \\   
9.2 & 22:01:52.060 & -59:56:41.59 \\
9.3 & 22:01:51.977 & -59:56:43.93 \\  
9.4 & 22:01:53.402 & -59:56:48.20 \\  
\enddata
\end{deluxetable}

Note that the two central de-magnified features 1.6 and 1.7 are completely outshone by the nearby BCGs G3 and G4 in the {\it HST} images.  These two features, however, are clearly detected in the [OII]$\lambda\lambda 3726.8,3729.2$ doublet emission in spectroscopic data taken with the Multi-Unit Spectroscopic Explorer (MUSE) Integral Field Unit (IFU) spectrograph.  For these features we use their coordinates as determined by \cite{Massey2}.

\subsection{\textit{Chandra}}\label{Chandra_data}
A3827 was observed with the \textit{Chandra} ACIS-S3 camera (ObsId 7920). We retrieve the archival data from the Chandra Data Archive and followed the standard data processing procedure by using CIAO v4.9 and CALDB v4.7.7.  We began with the level 1 event files and created an exposure-corrected image in the broad band (spanning 0.5$-$7 keV, with an effective energy of 2.3 keV) having an angular resolution of $0\farcs5$/pixel.  We located point sources with the CIAO task WAVEDECTECT using the default settings, from which a total of 15 point sources were identified. We removed these point sources and filled the removed regions with neighboring diffuse background emission using the CIAO task DMFILTH. 

\begin{figure}[tp!]
\centering
\includegraphics[width=\linewidth]{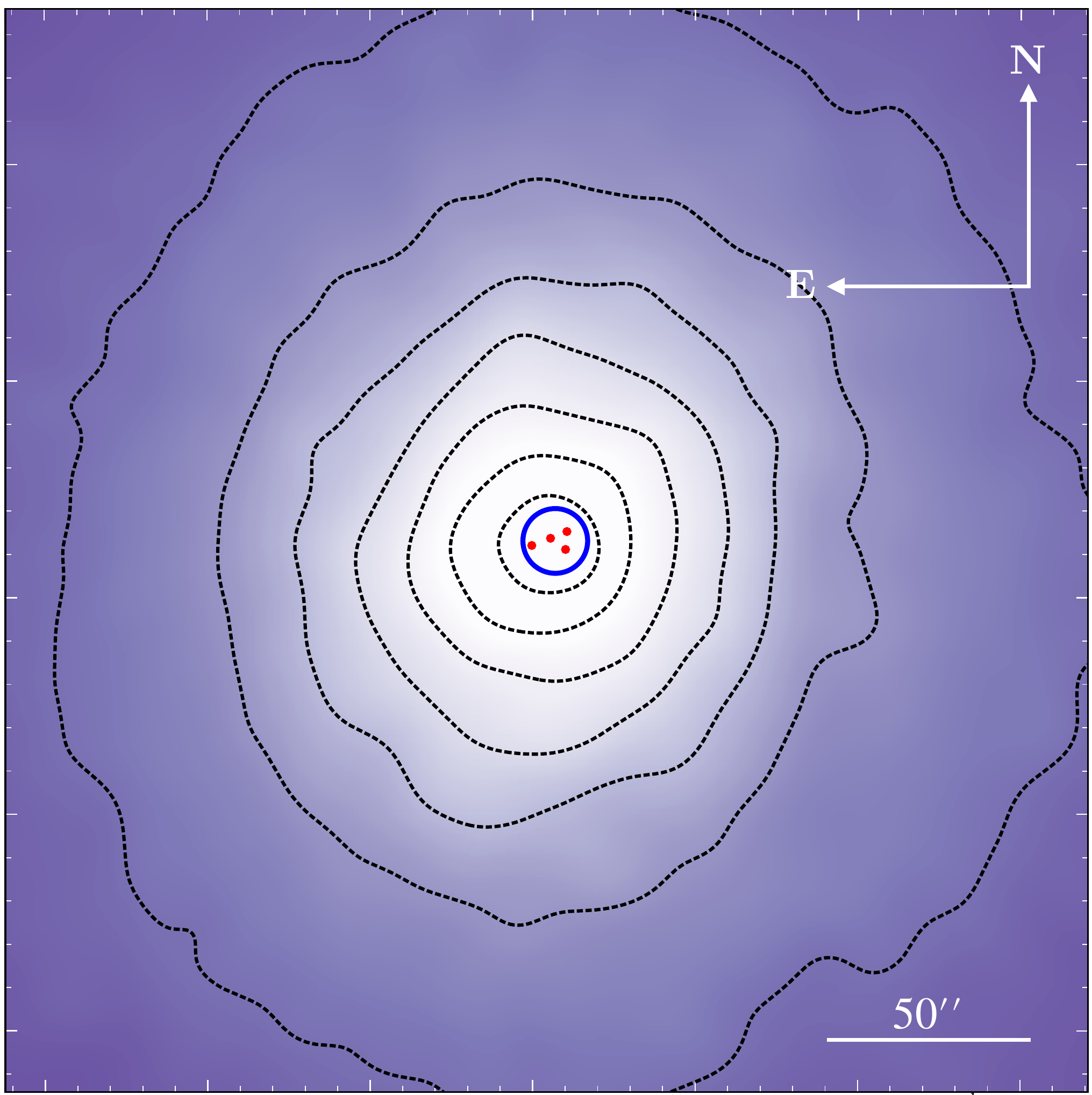}
\caption{\label{fig:x_ray_map} False-color image of the X-ray diffuse emission in the broad band (0.5$-$7 keV).  Point sources are removed, and the image is smoothed with a Gaussian kernel with a width of 15 pixels. The contours are spaced according to the square root of the emission intensity. Red circles indicate the locations of the four BCGs G1--G4; the blue circle indicates the approximate position of the Einstein ring formed from the background lensed galaxy at $z=1.24$.}
\end{figure}

Figure~\ref{fig:x_ray_map} shows the diffuse X-ray emission image of A3827 smoothed with a Gaussian kernel with a $\sigma$ of 15 pixels. Note that the smoothing is applied only for visual clarity. All subsequent analyses were conducted with the unsmoothed image.  The X-ray diffuse emission from A3827 is round and smooth. We find the ellipticity of the X-ray isophotes ($\epsilon_{\rm X}$) at all radii to be very small, as described next.

\subsubsection{Ellipticity of diffuse X-ray emission}\label{X-ray_ellipticity}
We estimate the ellipticity ($\epsilon_{\rm X}$) of the X-ray isophotes following the method in \cite{Buote1994}. We first locate the center of the diffuse X-ray emission by iteratively computing the first moments of data counts from the broadband emission,
\begin{equation}
\bar x  = \frac{1}{N}\sum_{i=1}^P n_i x_i \quad \textrm{and} \quad \bar y  = \frac{1}{N}\sum_{i=1}^P n_i y_i,
\label{eq:x_ray_center}
\end{equation}
where $n_i$ is the count value in pixel $i$ with Cartesian coordinates $(x_i,y_i)$, $P$ is the total number of pixels, and $N = \sum_{i=1}^P n_i$ (i.e., the total number of counts in $P$ pixels). We start by identifying the X-ray diffuse emission center by eye, and calculating $\bar x$ and $\bar y$ iteratively using Eq.~(\ref{eq:x_ray_center}) within a circular aperture $137\arcsec$ in radius (containing $\sim60\%$ of total counts) until the result changes by less than one pixel. With this method, we locate the center of the X-ray emission to be at RA=$330.47285\degree$ and Dec=$-59.946383\degree$, which we take to define the cluster center.

We then compute the ellipticity of the diffuse X-ray emission within circular apertures around the center using the iterative moment technique described in \cite{Buote1994}.  Specifically, the moments are computed according to
\begin{equation}
\mu_{mn} = \frac{1}{N}\sum_{i=1}^P n_i(x_i-\bar x)^m(y_i-\bar y)^n \quad  (m,n\le 2)
\label{eq:elli_moment}
\end{equation}
We then calculate the ellipticity 
\begin{equation}
\epsilon_X = 1-\frac{\Lambda_-}{\Lambda_+}
\label{eq:elli_epsilon}
\end{equation}
where $\Lambda_-$ and $\Lambda_+$ are the positive roots of the equation
\begin{equation}
(\mu_{20}-\Lambda^2)(\mu_{02}-\Lambda^2) = \mu_{11}^2
\label{eq:lambda}
\end{equation}  
for $\Lambda_- \le \Lambda_+$.

The $\epsilon_{\rm X}$ thus measured within different circular apertures with increasing radii up to $130\arcsec$ (13 times the size of the Einstein ring formed from the lensed galaxy at $z=1.24$) is listed in Table~\ref{table:X-ray_ellipticity}. In \cite{Buote1994}, subsequent iterative processes with refined elliptical apertures were conducted to increase the accuracy of the measured $\epsilon_{\rm X}$.  In the case of A3827, however, the ellipticity is consistently very small ($\epsilon_{\rm X}< 0.05$) for all radii. The ellipticity is especially small for smaller radii ($\mathrm{R}< 50\arcsec$), having values less than 0.025. 

\begin{deluxetable}{p{0.95cm} p{0.95cm} | p{0.95cm} }[tp!]
\tablecaption{\label{table:X-ray_ellipticity}Ellipticity ($\epsilon_{\rm X}$) of diffuse X-ray emission at different radii R.}
\tablehead{\colhead{R ($\arcsec$)} & \colhead{R (kpc)} & \colhead{$\epsilon_{\rm X}$}}
\startdata
10 & 18.3 & 0.023 \\
30 & 54.8 & 0.014 \\
50 & 91.4 & 0.008 \\
70 & 128.0 & 0.029 \\
90 & 164.5 & 0.041 \\
110 & 201.1 & 0.045 \\
130 & 237.6 & 0.046 \\
\enddata
\end{deluxetable}

\subsubsection{Gas mass}\label{X-ray_gas_mass}
Assuming the X-ray gas to have a spherically symmetric distribution, we estimate its mass by first fitting a double $\beta$ model to the radial profile in X-ray surface brightness so as to infer the radial profile in electron density \citep[e.g.][]{Ettori2000}. The best-fit parameters for the double $\beta$ model are $(r_1, \beta_1) = (40.0 \, \mathrm{kpc}, \, 3.0)$, and $(r_2, \beta_2) = (120.0 \, \mathrm{kpc}, \, 0.53)$. We then integrate the gas density along a given line of sight from $-R_{\mathrm{vir}}$ to $+R_{\mathrm{vir}}$ to obtain the projected mass as a function of radius, $R$, from the cluster center, as shown in Figure~\ref{fig:1D_mass}.  Note that within the Einstein ring at a radius of $\sim$$10\arcsec$ ($\sim$18 kpc) from the cluster center, the projected mass of the X-ray gas is an order of magnitude below that of the cluster member galaxies G1--G5 and the intracluster light summed together.

\section{Lens modeling}\label{lens_modeling}
With a good understanding of the projected light distribution of cluster member galaxies within the Einstein ring and the ICL, along with the mass distribution of the intracluster gas, we now construct both free-form and parametric lens models to reproduce the Einstein ring of A3827 using, respectively, the algorithms Weak and Strong Lensing Analysis Package $+$ \citep[WSLAP+;][]{Diego2005,Diego2007,Ponente2011,Sendra2014} and {\it glafic} \citep{glafic}. The center of all the lens models is at (RA=$330.470450$, DEC=$-59.945818$), the centroid of the intracluster X-ray gas. We use the multiply-lensed knots listed in Table~\ref{image-constraints} as constraints for both the free-form and parametric lens models. We do not use weak lensing constraints as we are not concerned with the gravitational potential in the outskirts of the cluster well beyond the Einstein ring. We construct all of the lens models in the framework of GR. In this section, we first present the free-form and parametric lens models that we derive separately, before providing an overall summary of their main features and their ability to reproduce the Einstein ring in A3827.   

\subsection{Free-from models by WSLAP+}\label{section:wslap}
\subsubsection{The algorithm}
WSLAP+ adopts a free-form philosophy whereby the lens plane is divided into a pixelated grid. Each pixel is represented as a Gaussian mass profile, having a full-width half maximum (FWHM) that can be varied to generate a multi-resolution grid or that is held constant to provide a uniform grid \citep{Diego2005}. The division of the lens plane into grid points allows us to divide the deflection field, $\alpha$, into the individual contributions to the deflection field from the pixel grid. A further improvement was implemented by including member galaxies of the cluster \citep{Sendra2014}, and for which the only free parameter is the scaling of the $M/L$ ratio for the member galaxies included in the model. This $M/L$ ratio and the Gaussian masses in the grid points are derived by minimizing a quadratic function; the minimum of this quadratic function is also the solution of a system of linear equations that describe the observed data, as described in more detail below.

Given the standard lens equation,
\begin{equation} \beta = \theta -
\alpha(\theta,\Sigma(\theta)) \, , \label{eq_lens} 
\end{equation} 
where $\theta$ is the observed position of the source, $\alpha$ is the deflection angle, $\Sigma(\theta)$ is the projected surface mass density of the cluster at the position $\theta$, and $\beta$ is the position of the background source. Both the strong lensing and weak lensing observables can be expressed in terms of derivatives of the lensing potential
\begin{equation} 
\psi(\theta) = \frac{4 G D_{l}D_{ls}}{c^2 D_{s}} \int d^2\theta'
\Sigma(\theta')ln(|\theta - \theta'|) \, , \label{eq_psi} 
\end{equation}
where $D_l$, $D_{ls}$ and $D_s$ are the angular diameter distances to the lens, from the lens to the source, and from the observer to the source, respectively. The unknowns of the lensing problem are in general the surface mass density (or mass in each grid cell) and the true positions of background lensed sources. The weak and strong lensing problem can be expressed as a system of linear equations that can be represented in a compact form \citep{Diego2007}, 
\begin{equation}
\Theta = \Gamma X, \label{eq_lens_system} 
\end{equation} 
where the measured strong and weak lensing observables are contained in the
array $\Theta$ of dimension $N_{\Theta }=2N_{SL} + 2N_{WL}$, the
unknown surface mass density and true positions of lensed sources are in the array $X$
of dimension $N_X=N_c + N_g + 2N_s$, and the matrix $\Gamma$ is known
(for a given grid configuration and fiducial galaxy deflection field, 
see below) and has dimension $N_{\Theta }\times N_X$.  $N_{SL}$ is the number
of strong lensing observables (each one contributing with two constraints,
$x$, and $y$), $N_{WL}$ is the number of weak lensing observables
(each one contributing with two constraints, $\gamma_1$, and $\gamma_2$), and $N_c$ is the number of grid cells that we use to divide
the field of view. $N_g$ is the number of deflection fields (from
cluster members) that we consider.  $N_s$ is the number of lensed background
sources (each contributes with two unknowns \citep{Sendra2014}, $\beta_x$, and $\beta_y$. 

The solution is found by minimizing a quadratic function that estimates the solution for Eq.~(\ref{eq_lens_system}). For this minimization we use a quadratic algorithm that is optimized for solutions such that the solution, $X$, must be positive \citep{Diego2005}. Imposing this constraint is particularly important to avoid the unphysical situation where the masses associated to the galaxies are
negative (that could, from the formal mathematical point of view, otherwise provide a reasonable solution to the system of linear Eq.~(\ref{eq_lens_system})). Furthermore, this constraint helps in regularizing the solution as it avoids large negative and positive contiguous fluctuations. In a previous study, we quantified via simulations how the addition of deflections from all of the main member galaxies helps improve the mass reconstruction with respect to our previous standard  non-parametric method \citep{Sendra2014}. Such perturbations cannot be recovered in grid-based reconstructions because the lensing information is too sparse to resolve member galaxies. 

WSLAP+ has been demonstrated to be able to provide robust lens models for not just virialized but also non-virialized galaxy clusters \citep[e.g.][]{Lam2014,Diego2018}, as well as a lens model that correctly predicted the re-appearance (both in time and location) of the first multiply-lensed supernova Refsdal \citep{Diego2016,Kelly2016}. 

\subsubsection{Model 1: With DM}\label{section:wslap_withDM}

We first generate a lens model in the context of $\Lambda$CDM by including a grid component representing the smooth mass distribution in the cluster, in addition to the mass sourced from G1--G5 and the ICL. We refer to this model as Model 1. The cluster member galaxies G1--G5 and the ICL are parameterized by scaling up the light distribution obtained from the best-fit \texttt{IMFIT} model (deconvolved with the PSF) in the F606W band, with a single $M/L$ ratio applied to all (i.e., $N_g = 1$). The optimization procedure determines the $M/L$ ratio that best reproduces the lensed images that serve as constraints for the lens model. The grid component is not constrained to follow the observed light distribution or any parametric function, reflecting the free-from nature of the algorithm. We use a regular grid with $8\times8=64$ cells (i.e., $N_c = 64$) for the lens model spanning a field of view of $0.8\arcmin \times 0.8\arcmin$. For this particular lens model construction, we do not explicitly define a component representing the intracluster X-ray gas (as we do in the lens models described later), but leave this component to be captured by the grid component.  

\begin{figure}[tp!]
\centering
\includegraphics[width=\linewidth]{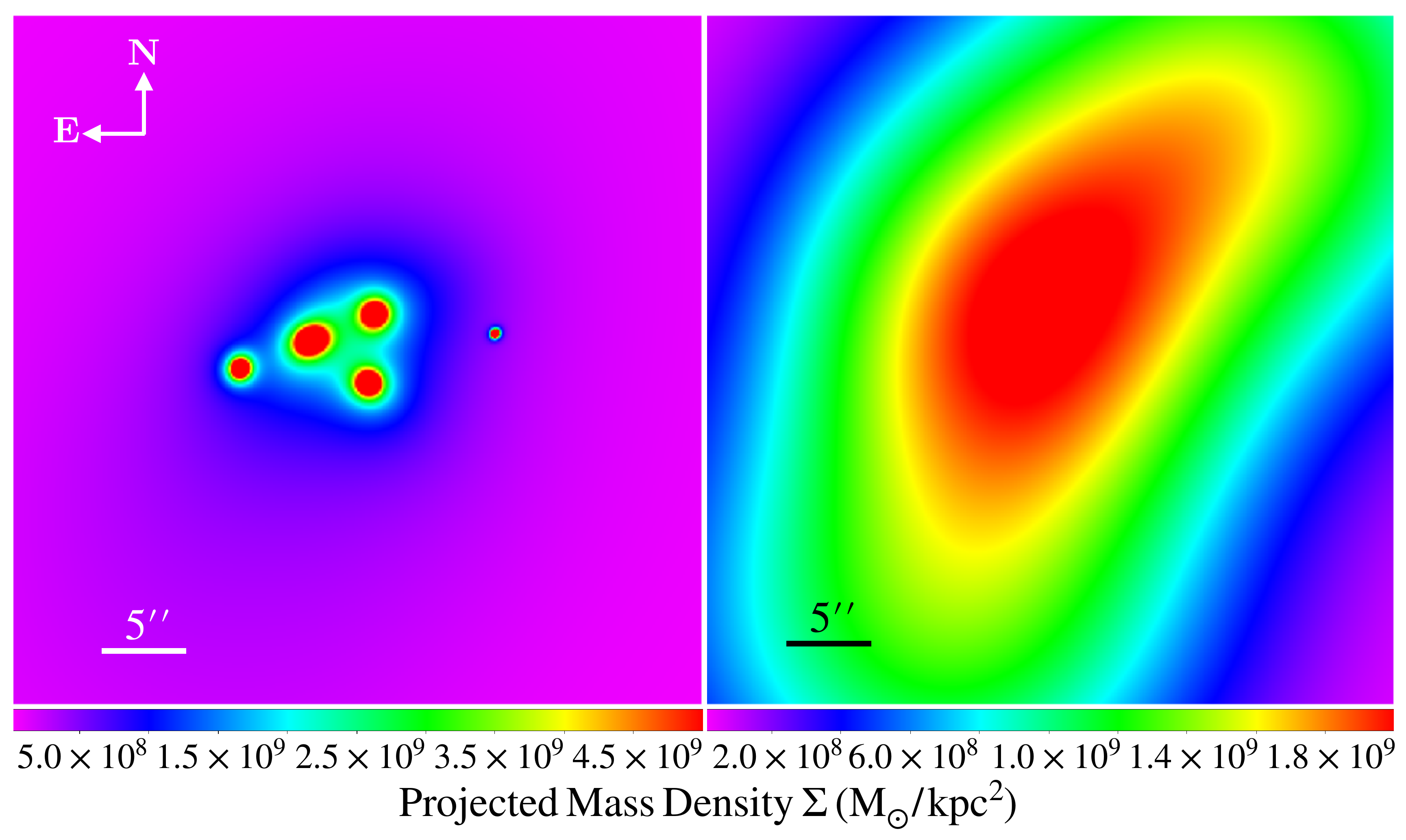}
\caption{\label{fig:model1_massmap}The galaxy (left) and grid (right) mass component of Model 1. The galaxy mass gives a $(M/L)_{\odot,\mathrm{F606W}}$ of $\sim9.7$, double the amount of stellar mass as obtained in the {\it Yggdraasil} model (see \S\ref{section:HST_data}). The grid component gives rise to a diffuse, cluster-scale mass distribution that dominates the total mass of the cluster. }
\end{figure}

The lens model thus derived has a total projected mass of $\sim$$2.7\times10^{12} \, M_\odot$ within the Einstein ring.  The best-fit mass-to-light ratio $(M/L)_{\odot,\mathrm{F606W}}$ for the galaxies G1$-$G5 and the ICL is $\sim$$9.7$, giving a total galaxy mass of $M_\mathrm{galaxy}\simeq2M_\star$ based on the stellar mass of galaxies inferred from the {\it Yggdrasil} model for the observed stellar light (see Table~\ref{table:stellar_mass}).  The grid component, which gives rise to a smooth cluster-scale mass distribution, dominates the total mass of the cluster.  In Figure \ref{fig:model1_massmap}, we show the projected mass density of the galaxies G1$-$G5 and the ICL (left panel) and that of the grid component (right panel) separately. As can be seen, the mass captured in the grid component does not have a circularly symmetric distribution unlike the intracluster X-ray gas, nor does it have a PA for its major axis related to that of any one of the galaxies G1 to G5 or their overall projected two-dimensional sky distribution. Instead, this dominant component has a PA similar to that of the ICL. We will further explore the shape of this diffuse mass distribution with parametric lens models later in \S\ref{section:glafic}. 

Using the lens model to delens multiply-lensed knots back to the source plane, we find a  root-mean-square (rms) dispersion in delensed positions from the mean delensed position for each set of multiply-lensed knots that averaged over all such sets is $\langle{\rm rms}_s\rangle=0\farcs29$. In Figure~\ref{fig:delensrelens}, we show the predicted appearance of the Einstein ring using A1 as the input (i.e., we delens image A1 back to the source plane, and lens it back to the image plane to obtain lens model predictions for its appearance at A2--A4).  The model-predicted appearances of A2--A4 closely resemble those observed, including the two central images that are outshone by G3 and G4 in the {\it HST} images but are clearly visible in [OII]$\lambda\lambda 3726.8,3729.2$ doublet emission in the MUSE data.
 

\subsubsection{Model 2: Without DM}\label{section_WSLAP_noDM}
Here we test the simplest possible lens model where all of the lensing mass is converted from visible light related to G1$-$G5 and the ICL with a fixed $M/L$ ratio (a condition we shall relax later in parametric model constructions). We explicitly omit the intracluster X-ray gas not only because its contribution to the gravitational potential potential within the Einstein ring is negligible (see Figure~\ref{fig:1D_mass}), but also to maximize the ability of the lens model to produce noncircular symmetry in the Einstein ring.  We refer to this model as Model 2. Procedurally, using WSLAP+, we exclude the grid component and simply scale up the light distribution of G1--G5 and the ICL obtained from the best-fit \texttt{IMFIT} model in the F606W band, as used in Model 1. The $(M/L)_{\odot,\mathrm{F606W}}$ ratio is the only free parameter of this model, and is optimized to be $15.7$. The total projected mass within the Einstein ring is $\sim$$1.9\times10^{12}\, M_\odot$, $\sim$30\% lower than that of Model 1.



\begin{figure*}[tp!]
\centering
\includegraphics[width=\linewidth]{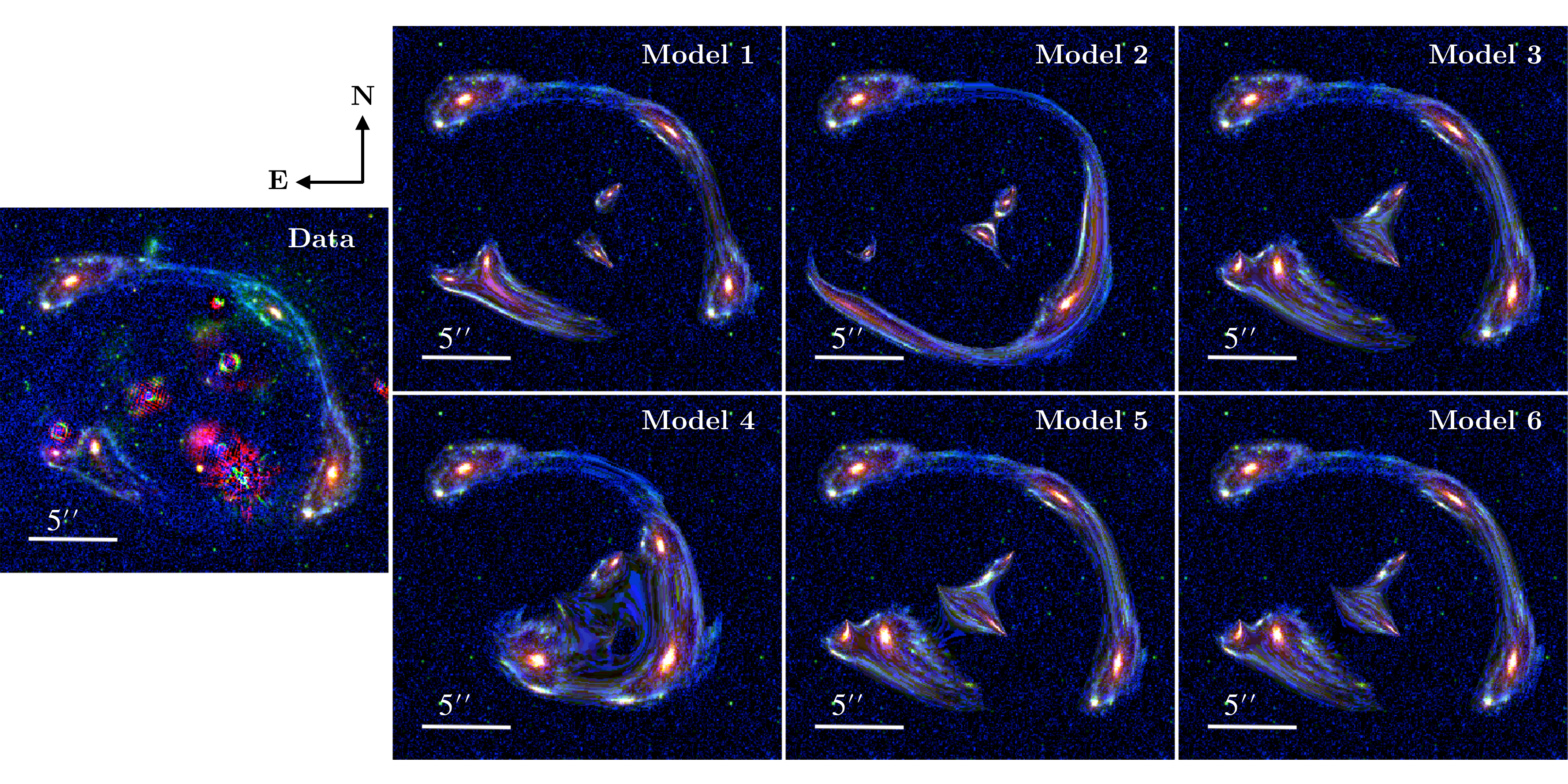}
\caption{\label{fig:delensrelens} Delens-relens results of all six models, using the image A1 as the input template. In the most left panel, we show a data image composed with residual images after subtracting the best-fit \texttt{IMFIT} models in each band from {\it HST} data. Models 1 and 2 are free-form models constructed by WSLAP+, as described in \S\ref{section:wslap}.  Models 3-6 are parametric models constructed by {\it glafic}, as described in \S\ref{section:glafic}. A summary of all six models is in \S\ref{section:model_summary}. Models 1, 3 and 6 are in the context of $\Lambda$CDM, where the lensing mass is dominated by a DM mass component that is not required to follow the observed light distribution. Model 2 is constructed by simply scaling up the light distribution (the best-fit \texttt{IMFIT} model in the F606W band). Models 4 and 5 are constructed by requiring the mass components in the lens model to follow observed light components, with a strict matching required in Model 4 and a more relaxed requirement in Model 5 induced by Gaussian priors.}
\end{figure*}

\begin{figure*}[tp!]
\centering
\includegraphics[width=16cm]{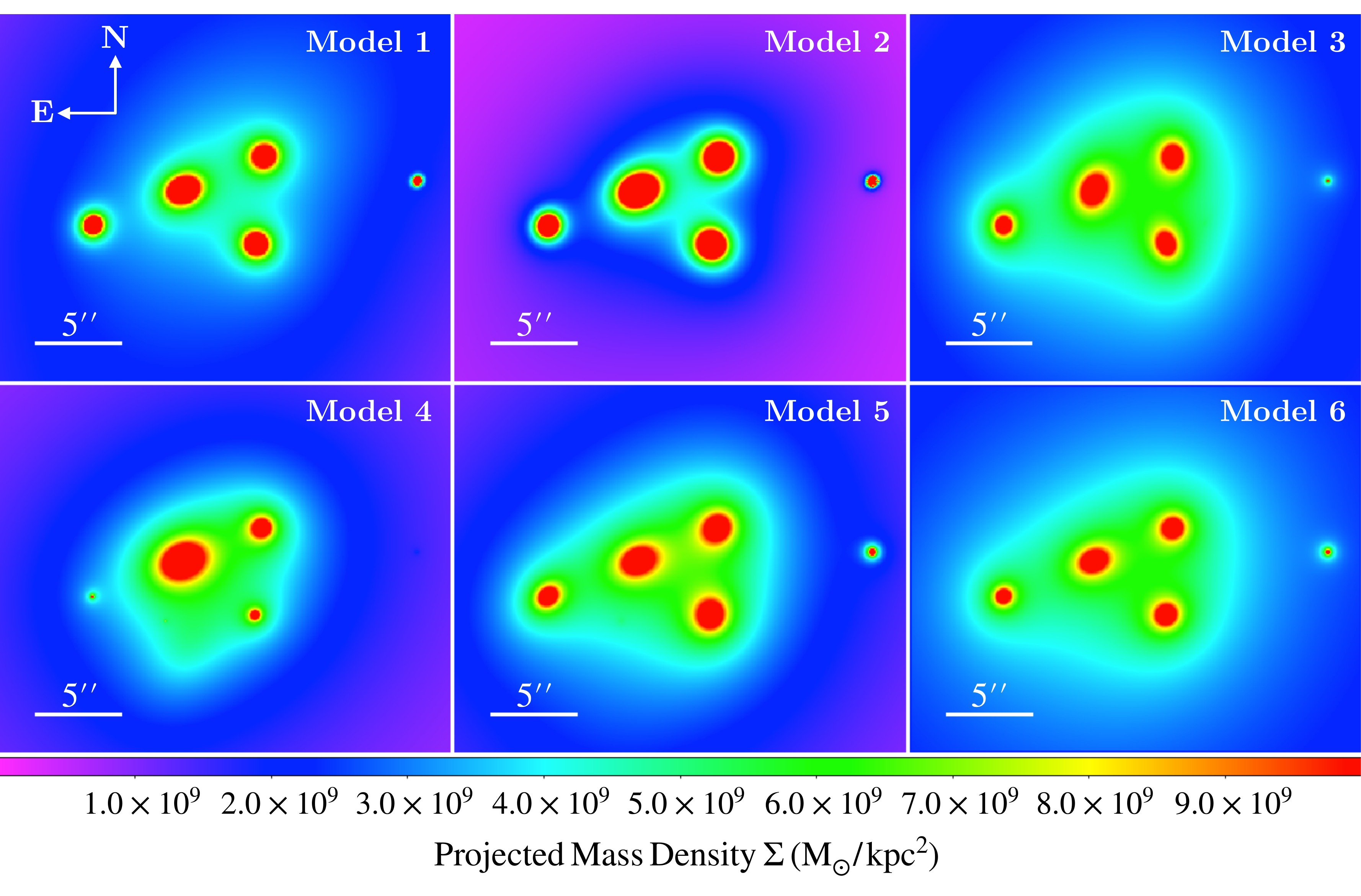}
\caption{\label{fig:massmap_allmodels}Maps of projected lensing mass in the cluster core of all six models. The corresponding delens-relens prediction by each model is shown in Figure~\ref{fig:delensrelens}, and a summary of all models is provided in \S\ref{section:model_summary}. }
\end{figure*}

Using this lens model to delens all sets of multiply-lensed knots back to the source plane, we find $\langle{\rm rms}_s\rangle=1\farcs494$, much worse than that of Model 1.  In Figure~\ref{fig:delensrelens}, we show the delens-relensed image predictions for Model 2 using image A1 as the input.  This lens model is able to produce an Einstein ring having roughly the observed size, but fails to predict the detailed morphology of images A2--A4. Specifically, when delensing and relensing A1, Model 2 completely fails to predict A2, predicts a bright extended arc to the north of A3 that does not match either the brightness or orientation of such a structure in the data, and fails to predict the observed bulge in A4. In Figure~\ref{fig:massmap_allmodels}, we show the total projected mass distribution of the two lens models. Owing to the lack of a dominating cluster-scale component as in Model 1, the lensing gravitational potential of Model 2 is much steeper than that of Model 1.

Model 2 demonstrates that a simple scaling of the visible light with a fixed $M/L$ ratio cannot generate a lensing mass capable of adequately reproducing the Einstein ring of A3827.
This finding motivates us to introduce spatially varying $M/L$ ratios for the visible light by changing the halo mass profiles of G1--G5, the ICL, and the intracluster X-ray gas, as we shall describe next.  As mentioned earlier, the introduction of spatially varying $M/L$ ratios also allows us to mimic alternative gravity theories that necessarily require the lensing gravitational potential to have a geometry tied to the projected two-dimensional sky distribution of visible matter.

\subsection{Parametric models by \textit{glafic}}\label{section:glafic}
To implement spatially varying $M/L$ ratios for the visible matter, we construct parametric lens model for A3827 using the code {\it glafic} \citep{glafic}. {\it glafic} is designed to compute parametric lens models for strong gravitational lensing by individual galaxies or entire galaxy clusters that lens point and/or extended background sources.  This algorithm can incorporate constraints from multiple image positions, relative brightness of lensed counterparts, and quasar image time delays. Here, we regard the identified multiply-lensed images listed in Table~\ref{image-constraints} as point sources in the source plane by using only the positions of these images as constraints.  Following \cite{Massey2}, we assign a positional uncertainty of $0\farcs5$ to image 1.6 and 1.7 (i.e., the nucleus of A6 and A7) and a smaller uncertainty of $0\farcs15$ to the rest of the images. 

\subsubsection{Model 3: With DM}\label{subsection:glafic_DM}
We start our parametric lens modeling by constructing a lens model in the context of $\Lambda$CDM that can be directly compared with those of \citet{Carrasco2010} and \citet{Massey1,Massey2} constructed also using parametric lensing algorithms in the same context.  We refer to this model as Model 3. We consider the lensing mass to comprise one cluster-scale DM halo and five galaxy-scale DM halos located at the positions of G1$-$G5; Neither the ICL nor the intracluster X-ray gas have specifically assigned components, a situation we shall remedy in later lens models. Specifically, we model the cluster-scale halo as a generalized NFW profile (gNFW) (e.g. \citealt{Jing2000}), whose radial profile is described as
\begin{equation}\label{gnfw_rho}
\rho(r)=\frac{\rho_s}{(r/r_s)^\alpha(1+r/r_s)^{3-\alpha}} \, ,
\end{equation}
and the two-dimensional convergence can be written as
\begin{equation}\label{gnfw_kappa}
\kappa(r) = \frac{1}{\Sigma_{crit}}\int^{\infty}_{-\infty}\rho(r)(\sqrt{r^2+z^2})dz.
\end{equation} 
The concentration parameter for the gNFW profile is defined as \begin{equation}
    c_{-2} = \frac{r_{vir}}{(2-\alpha)r_s} = \frac{c}{2-\alpha} \, ,
\end{equation}
where $c$ is the concentration parameter for the NFW profile with $\alpha=1$. We assign G1-G5 each with a pseudo-Jaffe mass profile \citep{Keeton2001}, the convergence of which is 
\begin{equation}\label{jaffe_kappa}
\kappa(r) = \frac{\theta_E}{2}(\frac{1}{\sqrt{r^2+r^2_{core}}}-\frac{1}{\sqrt{r^2+r^2_{trun}}}) \, ,
\end{equation} 
where $r_{core}$ is the core radius and $r_{trun}$ is the truncation radius.  The Pseudo-Jaffe profile generally provides a good fit to galaxy-scale lenses and allows great flexibility. For elliptical mass distributions, the above radius $r$ for both the gNFW and Pseudo-Jaffe profiles can be replaced with 
\begin{equation}
    r_\epsilon = \sqrt{\frac{x^2}{(1+\epsilon)^2}+\frac{y^2}{(1-\epsilon)^2}} \, ,
\end{equation}
where $\epsilon$ is the ellipticity and defined as $\epsilon = 1-b/a$ ($b$ is the semi-minor axis and $a$ is the semi-major axis). We also include external shear in this model, produced by a group of massive galaxies located $\sim$$2\arcmin$ away from the core of A3827 at a PA of $\sim$$120\degree$ (see Figure 1 in \citealt{Carrasco2010}). 

We allow all of the parameters of the cluster-scale gNFW halo to freely vary, while applying a flat prior on the position of the halo center requiring it to be within $5\arcsec$ radius from the X-ray emission center (RA$=330.47285\degree$, Dec$=-59.946383\degree$) (see \S\ref{X-ray_ellipticity}). The positions of galaxy-scale halos for G1--G5 are fixed at the light centroid of each galaxy as defined from the \texttt{IMFIT} fitting.  All the other parameters for the halos located at G1--G4 are allowed to freely vary; for G5, only the velocity dispersion and halo truncation radius are allowed to be free. The best-fit model is found by a standard $\chi^2$ minimization adopting a downhill simplex method. $\chi^2$ is evaluated in the source plane.  Readers can refer to \cite{glafic} for more details of model optimization. 

Reassuringly, we obtain a lens model similar to those obtained in the previous parametric lens modeling work of \citet{Carrasco2010} and \citet{Massey1,Massey2}, except for a smaller $\epsilon$ for the gNFW halo. The best-fit shear amplitude of $\gamma=0.02$ is reasonable given the presence of the nearby group of massive galaxies. In the absence of this external shear, a higher value of halo $\epsilon$ (as found by \citealt{Massey1}) is expected owing to degeneracy between halo $\epsilon$ and external shear \citep{Keeton1997}.  The best-fit parameters of Model 3 are listed in Table~\ref{table:best-fit_glafic_model3}; note that the $\epsilon$ and PAs of the mass halos for G1--G5 can be very different from the corresponding values for the visible light of these galaxies as listed in Table~\ref{table:BCG_ellipticity}.  This model has a reduced $\chi_{\nu}^2 = 1.02$, with a degree of freedom ($N_{\mathrm{dof}}$) of 29.  Delensed to the source plane, the rms positional dispersion among each set of multiply-lensed knots averaged over all such sets is $\langle{\rm rms}_s\rangle=0\farcs078$, exceeding the accuracy of Model 1. The better image prediction accuracy of Model 3 compared with Model 1 suggests that the mass distribution of A3827 can be well described by parametric profiles.  By contrast, the description of the cluster-scale halo in Model 1 is limited by the resolution of the grid cells. 

The Einstein ring predicted by Model 3 is shown in Figure \ref{fig:delensrelens}, whereas before we used image A1 as the input.  The model-predicted Einstein ring shows good agreement with that observed.  The projected surface mass density of this model is shown in Figure~\ref{fig:massmap_allmodels}. The total projected mass within the Einstein ring is $\sim4.5\times10^{12}\, M_\odot$, and is dominated by the cluster-scale halo (gNFW component).  This cluster-scale halo has a more slowly declining radial mass profile than the radial optical light profile, in agreement with Model 1 constructed using WSLAP+.  In the $\Lambda$CDM paradigm, the cluster-scale halo comprises predominantly DM.  The ICL is well aligned with this halo (difference in PA of $<15^\circ$), indicating that the ICL traces the large-scale gravitational potential.  As shown in \S\ref{sec:discussion_ICL}, the ICL has an $M/L \gtrsim60$, betraying the predominance of DM.

\begin{deluxetable*}{c | c c c c c c}[tp!]
\tablecaption{\label{table:best-fit_glafic_model3}Best-fit parameters of Model 3: {\it glafic} model in the context of $\Lambda$CDM. This model has $\chi_{\nu}^2 = 1.02$ ($N_{\mathrm{dof}}$=29). For gNFW profile, total mass is listed; for pseudo-Jaffe profile, the velocity dispersion is listed. Parameters in $[\,]$ are fixed values in the lens modeling.}
\tablehead{\colhead{} & \colhead{} & \colhead{$M_{\mathrm{total}}$ [$M_\odot$] (gNFW)} & \colhead{} & \colhead{} & \colhead{$c_{-2}$ (gNFW)} & \colhead{$\alpha$ (gNFW)} \\ \colhead{} & \colhead{Halo profile} & \colhead{or $\sigma_v$ [km/s] (Jaffe)} & \colhead{$\epsilon$} & \colhead{PA [$^\circ$]} & \colhead{or $r_{trun}$ [$\arcsec$] (Jaffe)} & \colhead{or $r_{core}$ [$\arcsec$] (Jaffe)}}
\startdata
Cluster & gNFW & $6.1\times10^{14}$ & 0.46 & 150.3 & 4.9 & 0.7 \\
G1 & pseudo-Jaffe & 231.4 & 0.11 & 353.5 & 39.1 & $\to$0 \\ 
G2 & pseudo-Jaffe & 286.5 & 0.23 & 157.3 & 40.1 & $\to$0 \\ 
G3 & pseudo-Jaffe & 252.2 & 0.15 & 0.0 & 40.5 & $\to$0 \\ 
G4 & pseudo-Jaffe & 238.7 & 0.21 & 198.8 & 38.0 & $\to$0 \\ 
G5 & pseudo-Jaffe & 79.0 & [0.0] & [0.0] & 2.6 & [$\to$0] \\ 
 & External shear & $\gamma=0.02$ & $\theta_\gamma=118.0$ & &  \\ 
\enddata
\end{deluxetable*}

\subsubsection{Model 4: Fixed Shape with Radially Varying $M/L$}\label{subsection:glafic_no_DM}
As demonstrated by Model 2 described in \S\ref{section_WSLAP_noDM}, a direct scaling of mass proportionally from light (i.e., a fixed $M/L$) for G1--G5 and the ICL cannot adequately reproduce the Einstein ring.  Here, taking advantage of parametric prescriptions to the lensing mass enabled by {\it glafic}, we construct a model where each of the mass components co-spatial with the visible matter (G1--G5, the ICL, and the intracluster X-ray gas) has core and truncation radii that can be freely varied, equivalent to applying a radially varying $M/L$ ratio, but has a shape defined by their light distribution.  We refer to this model as Model 4.  Note that there is no component representing a cluster-scale halo, which in $\Lambda$CDM would correspond to a DM halo.  Instead, with this parametrization, we mimic alternative gravity theories in which the geometry of the lensing gravitational potential is tied to the projected two-dimensional sky distribution of the visible matter.  In the framework of GR, such a radially varying $M/L$ is normally interpreted as different fractional mass in DM compared with visible matter as a function of radius.

Specifically, we construct Model 4 with seven pseudo-Jaffe mass halos: five for the galaxies G1--G5, one for the ICL, and one for the hot gas as seen in X-ray emission.  We fix the centers of these halos to be at the centroids of their corresponding light components, and the ellipticity and PAs to be their best-fit \texttt{IMFIT} parameters.  We also require the mass ratios of the seven halos to be in rough agreement with their flux ratios by assigning Gaussian priors on the ratios of their velocity dispersions. We allow the truncation and core radius of each halo to vary freely except for G5. The lens model constructed in this manner has a reduced $\chi^2_\nu = 143.30$ ($N_{\mathrm{dof}}$=39).  Delensed to the source plane, the rms positional dispersion among each set of multiply-lensed knots averaged over all such sets is $\langle{\rm rms}_s\rangle=1\farcs944$.  The best-fit model parameters are listed in Table~\ref{table:best-fit_glafic_model4}, and the projected surface mass density of this model is shown in Figure~\ref{fig:massmap_allmodels}.  The total projected mass within the Einstein ring is $\sim3.4\times10^{12}\, M_\odot$, comparable to that of Model 3 containing an additional component representing a cluster-scale halo (which dominates the mass even within the Einstein ring).
 
Using A1 as the input like before, we show the delens-relensed prediction of Model 4 for the Einstein ring in Figure~\ref{fig:delensrelens}.   The predicted appearance of the Einstein ring in this model resembles that in Model 2 (WSLAP+ model without a cluster-scale halo), thus confirming our previous conclusion that a projected mass distribution that strictly follows the projected shape of the visible matter will result in a lensing gravitational potential that is too steep to produce the observed lensed images.  Note that despite the Gaussian priors imposed on the relative mass ratios of the different mass halos, their best-fit mass ratios are very different from their relative brightness.

\begin{deluxetable*}{c | c c c c c c}[tp!]
\tablecaption{\label{table:best-fit_glafic_model4}Best-fit parameters of Model 4: {\it glafic} model with all mass halo shapes being fixed to the shape of their corresponding visible component. This model has $\chi^2_\nu = 143.30$ ($N_{\mathrm{dof}}$=39). }
\tablehead{\colhead{} & \colhead{Halo profile} & \colhead{$\sigma_v$ [km/s]} & \colhead{$\epsilon$} & \colhead{PA [$^\circ$]} & \colhead{$r_{trun}$ [$\arcsec$]} & \colhead{$r_{core}$ [$\arcsec$]}}
\startdata
G1 & pseudo-Jaffe & 80.1 & [0.05] & [139.07] & 39.2 & $\to$0 \\
G2 & pseudo-Jaffe & 369.9 & [0.22] & [113.15] & 177.0 & $\to$0 \\
G3 & pseudo-Jaffe & 239.2 & [0.04] & [125.76] & 114.5 & $\to$0 \\
G4 & pseudo-Jaffe & 160.3 & [0.06] & [121.56] & 76.4 & $\to$0 \\
G5 & pseudo-Jaffe & 24.8 & [0.0] & [0.0] & [1.00] & [$\to$0] \\
ICL & pseudo-Jaffe & 290.6 & [0.29] & [149.39] & 138.7 & 2.1 \\
Hot gas halo & pseudo-Jaffe & 27.7 & [0.0] & [0.0] & 300.0 & $\to$0 \\
 & External shear & $\gamma=0.13$ & $\theta_\gamma=136.9$ & &  \\ 
\enddata
\end{deluxetable*}

\subsubsection{Model 5: Free Shape with Radially Varying $M/L$}\label{subsection:glafic_free_varying_M/L}
Next, we relax the requirement that the mass halos assigned to the individual components of visible matter have the same shapes as their emitted light, and that their relative masses are in rough proportion to their relative brightnesses.
In this way, we can investigate which of the mass halos are required to have a different shape from, and/or contribute disproportionately in mass by contrast with, their emitted light.  In the framework of GR, any deviation in the shape of the mass distribution from that of the light distribution necessitates invoking DM. 

In Model 5, we therefore impose Gaussian priors on the ellipticities and PAs of all of the mass halos parametrized,  with a Gaussian width in $\epsilon$ of 0.2 and in PA of $15\degree$ centered at the observed $\epsilon$ and PA of each halo as listed in Table~\ref{table:BCG_ellipticity}. These generous priors allow the shape of the mass halos to differ significantly from the visible matter while preferentially sampling around the observed values. 
Similarly, we assigned Gaussian priors on the ratios of their velocity dispersions.  The best-fit parameters are listed in Table~\ref{table:best-fit_glafic_model5}.  This model has a reduced $\chi^2_\nu = 2.39$ ($N_{\mathrm{dof}}$=27), and after delensing to the source plane, it has an rms positional offset between multiply-lensed knots averaged over all such sets of $\langle{\rm rms}_s\rangle=0\farcs153$. We show the delens-relensed predictions of Model 5 for image A1 in Figure~\ref{fig:delensrelens}.  Although having a total projected mass  within the Einstein ring of $\sim4.3\times10^{12}\, M_\odot$ that is only slightly larger than that in Model 4, Model 5 is far superior to Model 4 in its ability to reproduce the observed Einstein ring, closely approaching the ability of Model 3 in this respect.  The projected surface mass density of Model 5 is shown in Figure~\ref{fig:massmap_allmodels}, and more closely resembles Model 3 rather than Model 4.

The primary differences between the parameterizations in Model 4 and Model 5 are for G1 and the ICL.  Both are optimized to have much higher halo ellipticities than their visible morphologies, albeit having similar PAs (difference $<10\degree$).  Furthermore, the ICL now makes a much larger contribution to -- disproportionate with its brightness -- and indeed dominates the total mass within the Einstein ring, acting like a massive elliptical halo in this model.  These important differences between the shapes of the best-fit mass halos and their visible matter, as well as their relative contributions in mass, disfavor the proposition that that the mass residing in those halos can be directly sourced from their visible matter.

\begin{deluxetable*}{c | c c c c c c}[tp!]
\tablecaption{\label{table:best-fit_glafic_model5}Best-fit parameters of Model 5: {\it glafic} model with Gaussian priors requiring a matched shape between mass halo shapes and their corresponding visible distribution shapes. This model has $\chi^2_\nu = 2.39$ ($N_{\mathrm{dof}}$=27).}
\tablehead{\colhead{} & \colhead{Halo profile} & \colhead{$\sigma_v$ [km/s]} & \colhead{$\epsilon$} & \colhead{PA [$^\circ$]} & \colhead{$r_{trun}$ [$\arcsec$]} & \colhead{$r_{core}$ [$\arcsec$]}}
\startdata
G1 & pseudo-Jaffe & 249.2 & 0.17 & 131.1 & 40.7 & $\to$0 \\
G2 & pseudo-Jaffe & 294.9 & 0.23 & 112.9 & 43.4 & $\to$0 \\
G3 & pseudo-Jaffe & 278.4 & 0.09 & 136.4 & 42.9 & $\to$0 \\
G4 & pseudo-Jaffe & 290.4 & 0.07 & 165.8 & 48.8 & $\to$0 \\
G5 & pseudo-Jaffe & 164.0 & [0.0] & [0.0] & [1.00] & [$\to$0] \\
ICL & pseudo-Jaffe & 442.2 & 0.58 & 158.4 & 120.1 & 10.0 \\ 
Hot gas halo & pseudo-Jaffe & 32.8 & 0.03 & 146.0 & 300.1 & 0.1 \\
 & External shear & $\gamma=0.07$ & $\theta_\gamma=146.9$ & &  \\ 
\enddata
\end{deluxetable*}

\begin{deluxetable*}{c | c c c c c c}[tp!]
\tablecaption{\label{table:best-fit_glafic_model6}Best-fit parameters of Model 6: Model 4 plus an extra free gNFW halo. This modle has $\chi^2_\nu = 1.40$ ($N_{\mathrm{dof}}$=32). }
\tablehead{\colhead{} & \colhead{} & \colhead{$M_{\mathrm{total}}$ [$M_\odot$] (gNFW)} & \colhead{} & \colhead{} & \colhead{$c_{-2}$ (gNFW)} & \colhead{$\alpha$ (gNFW)} \\ \colhead{} & \colhead{Halo profile} & \colhead{or $\sigma_v$ [km/s] (Jaffe)} & \colhead{$\epsilon$} & \colhead{PA [$^\circ$]} & \colhead{or $r_{trun}$ [$\arcsec$] (Jaffe)} & \colhead{or $r_{core}$ [$\arcsec$] (Jaffe)}}
\startdata
G1 & pseudo-Jaffe & 208.2 & [0.05] & [139.07] & 40.4 & $\to$0 \\
G2 & pseudo-Jaffe & 264.0 & [0.22] & [113.15] & 38.6 & $\to$0 \\
G3 & pseudo-Jaffe & 238.2 & [0.04] & [125.76] & 35.6 & $\to$0 \\
G4 & pseudo-Jaffe & 244.6 & [0.06] & [121.56] & 42.3 & $\to$0 \\
G5 & pseudo-Jaffe & 122.7 & [0.0] & [0.0] & [1.00] & [$\to$0] \\
ICL & pseudo-Jaffe & 50.5 & [0.29] & [149.39] & 101.1 & 2.1 \\
Hot gas halo & pseudo-Jaffe & 0.3 & [0.0] & [0.0] & 300.5 & 3.5 \\
Free halo & gNFW & $1.2\times10^{15}$ & 0.40 & 155.8 & 5.3 & 0.4 \\
 & External shear & $\gamma=0.01$ & $\theta_\gamma=117.8$ & &  \\ 
\enddata
\end{deluxetable*}

\subsubsection{Model 6: Free Shape and Radially Varying $M/L$, with DM}\label{subsection:glafic_free_varying_M/L_DM}
Finally, we construct Model 6, where we introduce a free gNFW halo back to the lens modeling, with its position and profile parameters free to vary while holding everything else the same as in Model 5.  In this way we obtain a good model with a reduced $\chi^2_\nu = 1.40$ ($N_{\mathrm{dof}}$=32), and after delensing to the source plane, it has an rms positional offset between multiply-lensed knots averaged over all such sets of $\langle{\rm rms}_s\rangle=0\farcs091$.  We list the best-fit parameters in Table~\ref{table:best-fit_glafic_model6}.  Note that the best-fit gNFW halo is required to have a significantly higher $\epsilon$, albeit the same PA, as the ICL, similar to that found in Model 5 for the best-fit mass halo assigned to the ICL.  We show the delens-relensed prediction of Model 6 for image A1 in Figure~\ref{fig:delensrelens}, and the projected mass density of this model in Figure~\ref{fig:massmap_allmodels}.  The total projected mass enclosed within the Einstein ring is $\sim4.5\times10^{12}\,M_\odot$, identical to that in Model 3. The lensing mass in Model 6 is dominated by the free gNFW halo, which is similar to the cluster-scale halo in Model 3. Model 6 confirms our previous conclusion that an extra component of mass that has a distribution different from the overall light distribution of G1--G5, the ICL, and the intracluster X-ray gas is required to accurately reproduce the Einstein ring of A3827.

\subsection{Summary of all models}\label{section:model_summary}

{\it Model 1:} a free-form model with DM. G1--G5 and the ICL are parameterized to follow their individual light distribution in the {\it HST} F606W band, along with a free-from grid component, constructed using WSLAP+ in the context of $\Lambda$CDM.  Within the Einstein ring, the inferred mass of the intracluster X-ray gas is negligible compared with that of stars and DM comprising G1--G5 and the ICL combined. The large-scale DM halo dominates the projected mass within the Einstein ring.  The ICL has a PA similar to that of the large-scale halo, indicating that the ICL traces the large-scale gravitational potential.  This model reproduces the Einstein ring well, and has an average positional dispersion between multiply-lensed knots delensed to the source plane of $\langle{\rm rms}_s\rangle=0\farcs29$. 

{\it Model 2:} same procedure as Model 1 but without a cluster-scale grid component, resulting in a different mass-to-light scaling for G1--G5 and the ICL.  This model fails to adequately produce the Einstein ring, demonstrating the need for a cluster-scale DM halo in the framework of GR.

{\it Model 3:} a preliminary parametric model constructed using  {\it glafic} to provide a direct comparison with previous parametric models by \citet{Carrasco2010} and \citet{Massey1,Massey2}.  G1--G5 are parameterized by five pseudo-Jaffe halos centered on these galaxies, to which is added a cluster-scale gNFW halo.  This model satisfactorily reproduces the Einstein ring, having $\chi_{\nu}^2 = 1.02$ ($N_{\mathrm{dof}}$=29) and an average positional dispersion between multiply-lensed knots delensed to the source plane of $\langle{\rm rms}_s\rangle=0\farcs078$.

{\it Model 4:} a parametric model that allows the $M/L$ ratio of all of the visible matter to freely vary with radius, while the $\epsilon$ and PA of each mass component are fixed to the observed values for their light. Thus this model mimics modified gravity theories that require the lensing gravitational mass to have a geometry tied to the projected two-dimensional sky distribution of the visible matter.  G1--G5, the ICL, and the hot intracluster medium are parameterized by seven pseudo-Jaffe halos centered on these objects.  We apply Gaussian priors on the relative masses of the different halos to be in rough agreement with the relative brightnesses of their associated light. This model has $\chi_{\nu}^2 = 143.20$ ($N_{\mathrm{dof}}$=39) and an average positional dispersion between multiply-lensed knots delensed to the source plane of $\langle{\rm rms}_s\rangle=1\farcs944$, providing a poor fit to the Einstein ring.

{\it Model 5:} same as Model 4 but without requiring a strict match between the shape of the mass halos and that of their associated light. Instead, we assign Gaussian priors to the $\epsilon$ and PA of all halos based on their observed values, as well as Gaussian priors to relative halo masses.  In this model, the halo ellipticities of G1 and the ICL are required to be much higher than that of their visible matter.  Furthermore, the halo of the ICL is required to dominate the projected mass within the Einstein ring, thus making a contribution that is disproportionate to its brightness.  This model has $\chi_{\nu}^2 = 2.39$ ($N_{\mathrm{dof}}$=27) and an average positional dispersion between multiply-lensed knots delensed to the source plane of $\langle{\rm rms}_s\rangle=0\farcs153$ in the source plane, and is able to reproduce the Einstein ring well.

{\it Model 6:} same as Model 4, but with an extra free gNFW halo.  The gNFW halo is found to have the same PA but higher $\epsilon$ than the ICL, and to dominate the projected mass within the Einstein ring.  This model has $\chi_{\nu}^2 = 1.40$ ($N_{\mathrm{dof}}$=32) and an average positional dispersion between multiply-lensed knots delensed to the source plane $\langle{\rm rms}_s\rangle=0\farcs091$, and is able to reproduce the Einstein ring well.

\section{Discussion}\label{sec:discussion}
\subsection{Thin-Lens Approximation}\label{discussion_MONDscale}



Models of gravitational lensing by galaxies and galaxy clusters usually adopt the thin-lens approach, whereby -- because distances between the background lensed object, lens, and the observer are much larger than the size of the lens -- the lensing mass is approximated as a thin sheet representing its projected mass in the sky plane.  In modified theories of gravity, however, the mass distribution along the line of sight can contribute significantly to the lensing potential, such that this approximation may not be valid.  Modeling the non-linear effects caused by the finite thickness of lensing objects using specific prescriptions from modified gravity theories is beyond the scope of this paper.  Here, we demonstrate that, based on the MOND theory, the gravitational acceleration at the positions of the model constraints in A3827 is in the Newtonian regime, and thus the thin-lens approximation should hold.

To estimate the Newtonian acceleration at the position of the Einstein ring ($R_\mathrm{E}\approx 10\arcsec$), we de-project the two-dimensional lensing mass obtained from Model 3 (best-fit parametric {\it glafic} model in the context of $\Lambda$CDM) through the inverse Abel transform, assuming spherical symmetry.  We obtain the three-dimensional density profile with
\begin{equation}
    \rho(r) = -\frac{1}{\pi}\int^{R_\mathrm{max}}_r \frac{1}{\sqrt{R^2-r^2}}\frac{d\bar{\Sigma} (R)}{dR}dR \, ,
\end{equation}
where $\bar{\Sigma (R)}$ is the mean two-dimensional lensing mass at distance $R$ from the center of lens modeling, averaged in annuli with a width of $0\farcs13$ ($\sim 240$ pc). The total mass within $R_\mathrm{E}$ is then calculated using 
\begin{equation}
    M(<R_\mathrm{E}) = 4\pi \int_0^{R_\mathrm{E}}r^2\rho(r)dr \, ,
\end{equation}
such that the Newtonian acceleration at the position of the Einstein ring is 
\begin{equation}
    a(R_\mathrm{E}) = \frac{GM(<R_\mathrm{E})}{R_\mathrm{E}^2} \, .
\end{equation}
Adopting $R_\mathrm{max}=1\arcmin$ and $R_\mathrm{E} = 10\arcsec$ (18.3 kpc at $z=0.099$), we estimate $M(<R_\mathrm{E})\approx 2\times 10^{12} M_\odot$ and $a(R_\mathrm{E}) \approx 9\times 10^{-10} \, \mathrm{m}\, \mathrm{s}^{-2}$. We have experimented with smaller and larger values of $R_\mathrm{max}$, which do not change the estimated value of $a(R_\mathrm{E})$ significantly, so that the Newtonian acceleration is reliably determined to be of the order of $10^{-9}\,\mathrm{m}\, \mathrm{s}^{-2}$ at the position of the lensing constraints. As $a(R_\mathrm{E}) > 10^{-10}\,\mathrm{m}\, \mathrm{s}^{-2}$, the gravitational acceleration is above the MOND scale, and hence a lens model constructed with thin-lens approximation should be appropriate even for MOND \citep{Mortlock2001}.

\subsection{A New Geometrical Challenge to Alternative Gravity Theories}\label{alternative_theories}

Testing modified gravity theories with gravitational lensing has previously been conducted in a variety of situations. These studies have provided mixed evidence for different classes of modified gravity formalism. We briefly describe below a few such studies to place our work in the appropriate context. 

On galactic scales, multiply-lensed quasars have been used to infer the lensing mass required in the MOND scheme \citep[e.g.,][]{Shan2008,Ferreras2008,Chiu2011}.  In the cases studied, \cite{Shan2008} and \cite{Chiu2011} found the MOND lensing mass to be in agreement with the estimated stellar mass of the lensing galaxy.  On the other hand, despite having overlapping objects with those studied by \cite{Chiu2011},  \cite{Ferreras2008} do not find such an agreement in some cases. As pointed out by \cite{Chiu2011}, the differences are likely caused by minor differences in the detailed lensing formalism adopted for calculating the expected gravitational potential with MOND. In the theory of emergent gravity, like in MOND, gravity is sourced only from visible matter, but with apparent DM as an additional mass component \citep{Verlinde2017}. To a good approximation, the theory of Emergent Gravity can be tested by comparing the amount of DM required in a lensing system according to GR and the amount of apparent DM in emergent gravity as computed from the visible matter \citep[e.g., see][]{Brouwer2017,Ettori2017}. In this way, \cite{Brouwer2017} computed the mean mass density profile of $\sim$30,000 galaxies constrained by weak lensing data, and found that their DM mass as inferred from GR agrees with their apparent DM mass predicted by emergent gravity based on the stellar light and prescriptions for other matter (gas and satellite galaxies) in these galaxies.

On galaxy cluster scales, modified gravity theories have been confronted with more severe challenges.  Based on optical and X-ray data for three galaxy clusters, \cite{Takahashi2007} concluded that the shear profile predicted by MOND is too shallow to explain the lensing signal.  By combining strong and weak lensing signals in six galaxy clusters, \cite{Natarajan2008} found that the amount of invisible matter cannot be explained by MOND even with a component of 2\,eV neutrinos. Computing the radial mass profile of two massive galaxy clusters based on hydrostatic equilibrium with physical parameters inferred from the intracluster X-ray gas, \cite{Ettori2017} found a factor of two to three discrepancy in the inferred mass of DM compared with the apparent mass of DM predicted by emergent gravity at the inner region of these clusters, although they found good agreement at the cluster outskirts.

All of the aforementioned studies rely upon comparing the enclosed or radial profile in the lensing mass deduced by a given theory of gravity with that of the visible matter.  In our work, we describe a fundamentally different test, where we compare the two-dimensional shape of the lensing mass as projected onto the sky with that of the visible matter.  Although we do not specifically compute the lensing mass in the framework of alternative gravity theories, we mimic these theories by allowing the $M/L$ ratio of the mass halos assigned to each component of visible matter (cluster member galaxies, the intracluster stellar light, and the intracluster X-ray gas) to vary with radius.  In doing so, we find that the Einstein ring of A3827 cannot be reproduced by lens models in which the matter distribution follows the shapes of the visible matter components, as is implicitly required in alternative theories of gravity such as MOND and emergent gravity.


\begin{figure*}[tp!]
\centering
\includegraphics[width=\linewidth]{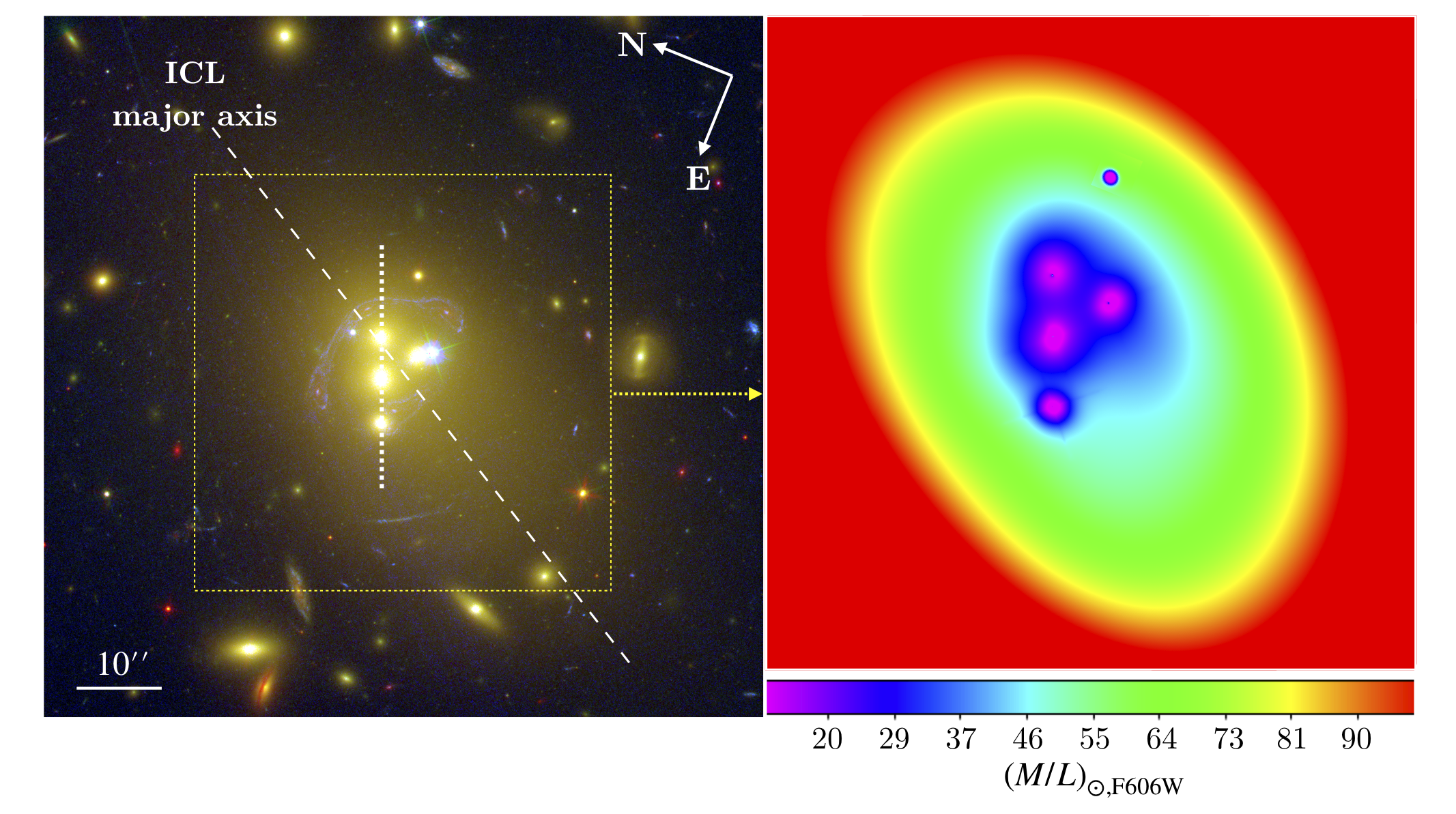}
\caption{\label{fig:hst_show_offset}{\it Left: }a color image of the A3827 cluster, composed with {\it HST} images from filters F336W (blue), F606W and F814W (green) and F160W (red). There is extended intra-cluster light (ICL) in the cluster core, with a major axis (shown in dashed line) significantly offset from the alignment of the three most massive BCGs (shown in dotted line). As we have shown through the lens modeling process, the ICL traces the large-scale gravitational potential induced by an elliptical, smooth mass distribution in this cluster that is primary in the form of DM. {\it Right: }the mass-to-light ratio map using the lensing mass obtained from the best-fit {\it glafic} Model 3. Towards the centers of individual galaxies G1--G4, the $M/L$ ratio value $(M/L)_{\odot,\mathrm{F606W}}$ is around 10, typical for massive elliptical galaxies.}
\end{figure*}

\subsection{Necessity for DM}\label{sec:discussion_ICL}

The shape ($\epsilon$ and PA) of the cluster-scale halo in the free-form lens model (Model 1) is very similar to that in the parametric lens models (Models 3 and 6).  This agreement is not guaranteed given the freedom of the pixel grid in free-form lens modeling compared with a fixed radial profile adopted in parametric lens modeling, and provides confidence in the parameters inferred for the cluster-scale halo necessary for producing the Einstein ring.

As pointed out above, the ICL, which extends well beyond the Einstein ring as shown in Figure~\ref{fig:hst_show_offset}, has a major axis that is closely oriented with that of the cluster-scale halo, which, when included in the lens modeling, dominates the total mass of the cluster. This agreement in their PAs indicates that the ICL traces the gravitational potential defined by the large-scale halo, which is to be expected if stars that produce the ICL were tidally stripped from cluster member galaxies. Such an agreement is also demonstrated by \cite{Montes2019} for massive lensing clusters in the {\it Hubble} Frontier Fields.  On the other hand, the intracluster X-ray gas is circular as projected onto the sky.  This circular morphology of hot intracluster medium is not surprising due to its collisional nature, in contrast with collisionless DM particles and intracluster stars \citep[e.g.][]{Lee2003}. 

If the ICL traces the large-scale lensing gravitational potential, can it  contribute entirely or predominantly to this lensing mass? As shown in Figure~\ref{fig:hst_show_offset} (right panel), in the absence of DM, the ICL needs an extremely high $(M/L)_{\odot,\mathrm{F606W}}$ of $>60$ to account for the extra lensing mass required to produce the detailed Einstein ring. This $M/L$ ratio is an order of magnitude higher than that of the galaxies G1--G5, and seems implausible given the observed similarity in the colors of the ICL and these galaxies. The ICL therefore cannot be the primary source that contributes to the dominate, large-scale lensing mass. 

\section{Summary and Conclusions}\label{sec:conclusion}
We have posed the following question to test alternative gravity theories where gravity is sourced only from visible matter: can a mass distribution that strictly follows the visible matter distribution produce a particularly detailed Einstein ring in the galaxy cluster A3827? In doing so, we introduce a new test for alternative theories of gravity based upon the geometry of gravitational lensing, requiring not just the predicted lensing mass enclosed within a given radius to agree with the enclosed mass of visible matter, but requiring also the predicted two-dimensional sky distribution of the lensing mass to be tied to that of the visible matter.

To address this question, we first determined the two-dimensional distribution of visible matter in and around the Einstein ring of A3827 -- four dominant cluster member galaxies along with another member galaxy near the cluster center, ICL extending far beyond the Einstein ring, and X-ray emitting intracluster medium -- by fitting appropriate analytical functions to their light distributions ($\S\ref{data}$).  We then used both free-form (WSLAP+; Models 1--2) and parametric ({\it glafic}; Models 3--6) lensing algorithms to derive the lensing mass ($\S\ref{lens_modeling}$) as constrained by 39 knots in the Einstein ring, with these knots corresponding to 9 distinct features in the background lensed galaxy. 

We find that when we require a strict match between the shape ($\epsilon$ and PA) of the lensing mass and that of visible matter, we cannot reproduce the Einstein ring of A3827 for either a fixed $M/L$ (Model 2, $\S\ref{section_WSLAP_noDM}$) or radially varying $M/L$ ratios (Model 4, $\S\ref{subsection:glafic_no_DM}$). We then relax this strict requirement and allow their shapes also to vary, and find that the lensing masses of one of the dominant cluster member galaxies and the ICL are required to have much higher ellipticities than their corresponding light distributions (Model 5, $\S\ref{subsection:glafic_free_varying_M/L}$). Furthermore, in this model, the ICL is required to have an $M/L$ ratio many times higher than that of the dominant cluster member galaxies despite all having similar colors. Instead, by allowing a smooth, freely oriented DM halo in the lens models, we can accurately reproduce the Einstein ring while fixing the shapes of lensing mass components to the shapes of their corresponding light distributions, for either a fixed $M/L$ ratio (Model 1, $\S\ref{section:wslap_withDM}$) or radially varying $M/L$ ratios (Model 3, $\S\ref{subsection:glafic_DM}$; Model 6, $\S\ref{subsection:glafic_free_varying_M/L_DM}$) for the visible components.

Our work therefore shows that a mass distribution strictly tied to the observed distribution of visible matter in A3827 cannot adequately produce its Einstein ring.  Instead, a dominant cluster-scale mass component that has no visible counterpart is required. Throughout our work, we have not used specific prescriptions from alternative gravity theories to create a lens model, and all our lens models described above are derived in the framework of GR. However, we have demonstrated that at the location of lensing constraints, the gravitational acceleration is above the MOND scale (\S\ref{discussion_MONDscale}). Furthermore, we have constructed a model that mimics a generic version of modified gravity theory by fixing the shapes of lensing mass components to the shapes of their corresponding observed light, while allowing the $M/L$ ratios to vary radially (Model 4, \S\ref{subsection:glafic_no_DM}), and have demonstrated that such a model has difficulty reproducing the Einstein ring of A3827.  

The challenge therefore for alternative gravity theories is to create a lens model for the Einstein ring of A3827 that agrees not only with the inferred mass of visible matter in this cluster -- a challenge that both emergent gravity and MOND sometimes successfully pass in lens modeling ($\S\ref{alternative_theories}$) -- but that also agrees with the geometric distribution of the visible matter.

\acknowledgements
The authors thank Richard Massey for useful discussions, and the anonymous referee for constructive suggestions to improve this paper. TJB was supported by a Visiting Research Professor Scheme from the University of Hong Kong, during which major parts of this work were conducted. JL acknowledges support from the Research Grants Council of Hong Kong through grants 17319316 and 17304519 for the conduct and completion of this work. JMD acknowledges the support of project AYA2015-64508-P (MCIU/AEI/MINECO/FEDER, UE) funded by the Ministerio de Economia y Competitividad and project PGC2018-101814-B-100 (MCIU/AEI/MINECO/FEDER, UE) Ministerio de Ciencia, Investigaci\'on y Universidades. This work was supported in part by World Premier International Research Center Initiative (WPI Initiative), MEXT, Japan, and JSPS KAKENHI Grant Numbers JP15H05892 and JP18K03693. The scientific results reported in this work are based in part on data obtained from the {\it Chandra} Data Archive. 
\clearpage
\appendix
\renewcommand*{\thetable}{\Alph{section}\arabic{table}}
\renewcommand*{\thefigure}{\Alph{section}\arabic{figure}}

\section{Information of multiply-lensed knots}
\setcounter{figure}{0}
We plot in Figure~\ref{fig:image_ID} the locations of multiply-lensed knots used as constraints for lens modeling. Coordinates of these knots are listed in Table \ref{image-constraints}.

\begin{figure*}[tp!]
\centering
\includegraphics[width=14cm]{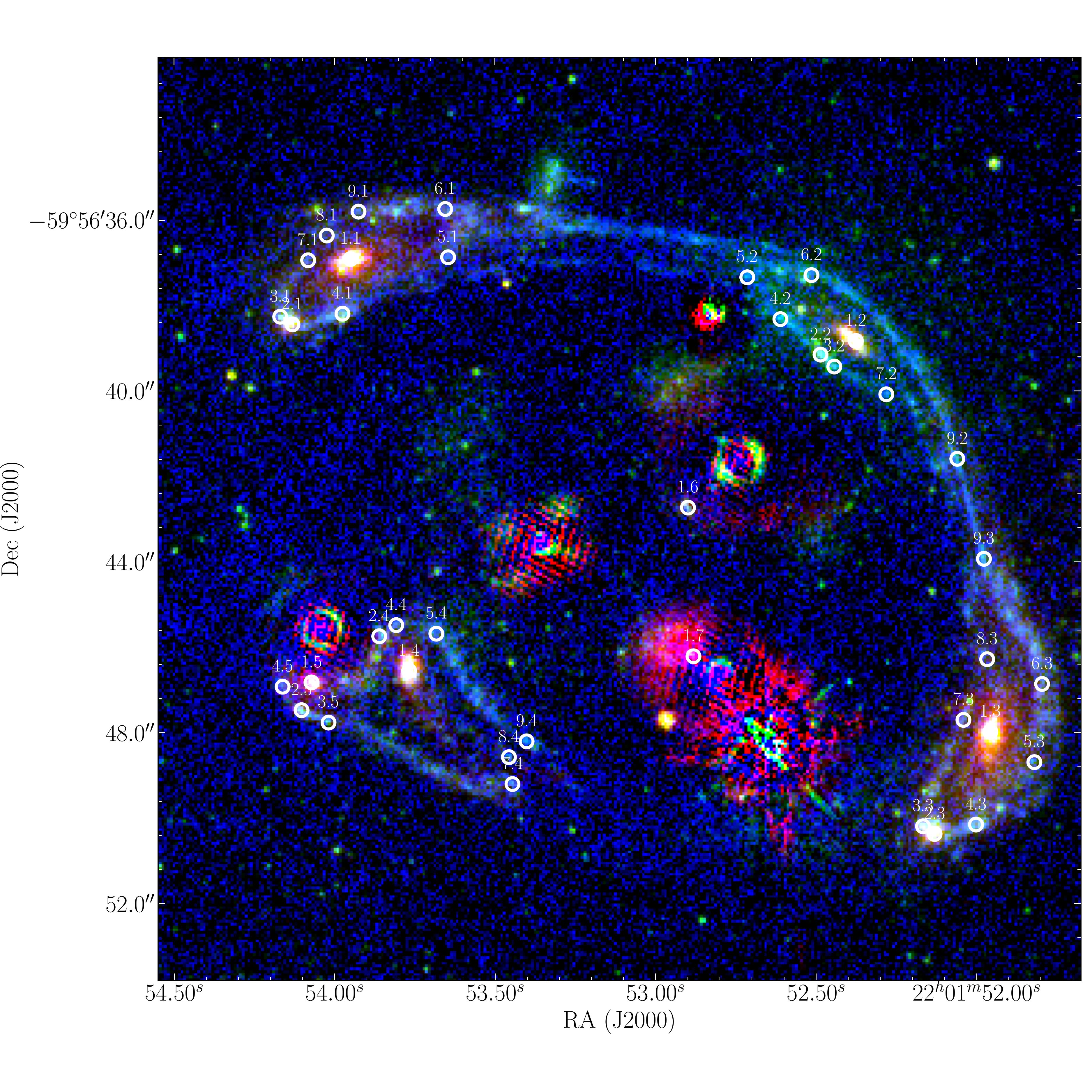}
\caption{\label{fig:image_ID} Locations of the 39 multiply-lensed knots used as constraints for constructing both WSLAP+ and {\it glafic} models. The coordinates of these knots are listed in Table~\ref{image-constraints}.}
\end{figure*}


\twocolumngrid
\bibliographystyle{aasjournal}
\bibliography{A3827_paper_ApJ}


\end{document}